\documentclass[reqno,10pt]{article}
\usepackage[bookmarks=true,linkcolor=red,citecolor=blue,urlcolor=green,colorlinks=true]{hyperref}
\usepackage{amsmath}
\usepackage{mathrsfs}
\usepackage{amssymb}
\usepackage{enumerate}
\usepackage{setspace}
\usepackage{url}
\usepackage{amsmath}
\usepackage{amsfonts}
\usepackage{amssymb}
\usepackage{lscape}
\usepackage{booktabs}
\usepackage{bigstrut}
\usepackage{listings}
\usepackage{qcircuit}
\usepackage{tikz}
\usetikzlibrary{arrows.meta}

\usepackage{algorithm}
\usepackage[noend]{algorithmic}
\usepackage{multirow,rotating}
\usepackage{amssymb,amsmath}
\usepackage{booktabs}
\usepackage{graphicx}
\usepackage{makeidx}  
\usepackage{multirow}
\usepackage{subcaption}
\usepackage{enumerate}
\usepackage{verbatim}
\usepackage{theorem}
\usepackage{caption}
\usepackage{ifpdf}
\usepackage{mathtools}
\usepackage{array}

 {{\nopagebreak\hspace*{\fill}$\Box$\par\vspace{12pt}}}

\newcommand{\R}{\mathbb{R}}

\DeclarePairedDelimiterX\braket[2]{\langle}{\rangle}{#1 \delimsize\vert #2}
\DeclarePairedDelimiterX\dotp[2]{\langle}{\rangle}{#1, #2}

\allowdisplaybreaks
\begin{document}

\title{Optimal qubit assignment and routing via integer programming}

\author{Giacomo Nannicini\thanks{IBM Quantum, IBM T.J. Watson Research Center, Yorktown Heights, NY, USA. \texttt{nannicini@us.ibm.com}} \and 
Lev S.~Bishop\thanks{IBM Quantum, IBM T.J. Watson Research Center, Yorktown Heights, NY, USA. \texttt{lsbishop@us.ibm.com}} \and 
Oktay Gunluk\thanks{Operations Research and Information Engineering, Cornell University, Ithaca, NY, USA. \texttt{ong5@cornell.edu} } \and
Petar Jurcevic\thanks{IBM Quantum, IBM T.J. Watson Research Center, Yorktown Heights, NY, USA. \texttt{petar.jurcevic@ibm.com}} 
}

\date{}
\maketitle

\begin{abstract}
  We consider the problem of mapping a logical quantum circuit onto a given hardware with limited two-qubit connectivity. We model this problem as an integer linear program, using a network flow formulation with binary variables that includes the initial allocation of qubits and their routing. We consider several cost functions: an approximation of the fidelity of the circuit, its total depth, and a measure of cross-talk, all of which can be incorporated in the model. Numerical experiments on synthetic data and different hardware topologies indicate that the error rate and depth can be optimized simultaneously without significant loss. We test our algorithm on a large number of quantum volume circuits, optimizing for error rate and depth; our algorithm significantly reduces the number of CNOTs compared to Qiskit's default transpiler SABRE \cite{li2019tackling}, and produces circuits that, when executed on hardware, exhibit higher fidelity.
\end{abstract}

\section{Introduction}
\label{sec:introduction}
Quantum computers promise to solve certain computational tasks asymptotically faster than any classical computer, and their engineering advances at rapid pace. Existing quantum computers have limitations, and the implementation of a quantum circuit on physical hardware must contend with the limitations of the device. Some of the prevalent technologies for the circuit model (e.g., superconducting qubits) only allow interactions between qubits that are physically adjacent on a chip. Furthermore, due to decoherence and numerous sources of noise, qubits degrade over time, favoring shallow circuits over deep circuits, and gates are imperfect, favoring circuits with a small gate count. As a consequence, implementing a logical quantum circuit on a specific hardware is a nontrivial task: there is an infinite family of circuits that approximately implement the same target unitary, and to choose among this family, all the previous considerations should be taken into account.

To reduce the complexity of the problem and bring it down to a manageable form, we make a simplifying assumption: we assume that we are given a circuit already decomposed as a sequence of one- and two-qubit gates that are part of the gate set available on the hardware, so that the only remaining task is to map logical qubits to physical qubits, and to ensure that the gates are applied in the correct order. Our algorithm to solve this problem is guided by some simplified models of fidelity of the circuit. 

With these assumptions, we are able to give a general and flexible integer programming formulation for the problem of mapping a logical circuit to the hardware. The hardware is described by an undirected graph representing qubits and their connectivity, and we allow each edge in the graph to have a different weight representing the fidelity of a CNOT applied on that edge. The logical circuit is described by a sequence of two-qubit gates (this is w.l.o.g., as we can merge one-qubit gates with adjacent two-qubit gates), and we can insert SWAP gates as necessary. We do not restrict the topology of the hardware, but the number of variables and constraints of the resulting integer program depends on the number of edges in the hardware graph; this is advantageous, because current superconducting qubit hardware implementations of the circuit model have extremely sparse topologies. Our formulation is akin to a network flow problem with design variables, whereby logical qubits are assigned to an initial location in a time-expanded version of the hardware graph, and they flow through the time-expanded network as a consequence of qubit-swaps. An important feature of our formulation is that it distinguishes between the case in which a SWAP is inserted and needs to be implemented with three CNOTs, and the case in which a SWAP can be merged with an adjacent two-qubit gate on the same pair of qubits, thereby requiring fewer additional CNOTs in comparison. This formulation was instrumental in demonstrating quantum volume \cite{cross2019validating} 64 on an IBM device \cite{jurcevic2021demonstration}; this was mentioned and briefly described in \cite{jurcevic2021demonstration}, here we give for the first time a full description of the mathematical model, of the methodology used to perform to optimization, and a comparison with existing algorithms.

To show the flexibility of our model, we consider three possible objective functions for the problem. The first objective function is to minimize the error rate of the circuit; the error rate is computed based on a first-order approximation that takes into account the fidelity of each two-qubit gate (depending on how many CNOTs are used in its decomposition), and the error rate of each CNOT, where for simplicity that errors are modeled as independent Bernoulli trials. The second objective function is to minimize the depth of the circuit, i.e., the number of layers such that at any layer, each qubit is involved in at most one gate. The third objective function is to minimize a measure of cross-talk, which we model via sets of hardware edges that should not be used simultaneously, if possible. Numerical experiments on quantum volume circuits (which are specifically designed to stress the capabilities of noisy quantum computers, by enforcing all-to-all connectivity) indicate that it is essentially possible to optimize the error rate and the depth at the same time, but minimizing cross-talk is in conflict with the two other objectives; thus, if cross-talk is a significant source of noise, one must choose which objectives to prioritize.

To optimize the qubit assignment and routing problem (also called {\em qubit allocation}) in practice, we choose to optimize the error rate first, followed by depth. We run several experiments applying the proposed methodology to 6-qubit and 8-qubit quantum volume circuits, on different hardware topologies. We compare results obtained solving our model using an off-the-shelf integer programming solver with the heuristic algorithm SABRE \cite{li2019tackling}; all 6-qubit circuits are solved to global optimality, while for 8-qubit circuits we apply our method as a heuristic, running it for limited time. Results indicate that our model can significantly reduce the CNOT count and depth of the circuit. This, in turn, has an impact on the fidelity of the circuit when executed on IBM's quantum hardware, which we verify by running the compiled 6-qubit quantum volume circuits, and recording the observed heavy output probability (HOP) \cite{aaronson2016complexity,cross2019validating}: the circuits globally optimized with our methodology record higher HOP, closer to the ideal value $(1+\ln 2)/2$, while those optimized with SABRE are further from it. Our experiments additionally indicate that the biggest gain of our algorithm, as compared to SABRE, comes from the optimal routing: the impact of choosing a suboptimal qubit assignment is less severe than choosing a suboptimal routing. However, optimizing both at the same time (as our model is capable of doing) yields significant benefits on the tested circuits. Experimental data shows that the proxy objective functions employed (i.e., error rate and depth) correlate with the recorded HOP; the total duration of the circuit would be an even more accurate predictor of the HOP, but because it depends on many factors, our integer linear programming model (which represents a stylized version of the quantum circuit) cannot optimize it directly. On average, the integer linear programming formulation reduces the CNOT count and duration by $\approx 10\%$ and $\approx 20\%$ compared to SABRE, even when used only to guide a heuristic search. To prove the usefulness of our model beyond quantum volume circuits, we compare it to SABRE on a set of toy Hamiltonian evolution circuits taken from \cite{bravyi2021clifford}, which employ several different topologies; on these circuits, which are all Clifford, the integer linear programming formulation reduces the CNOT count by $\approx 11\%$ on average.

\paragraph{Literature review.} The problem of mapping logical circuits to hardware with limited connectivity has attracted a lot of attention in the past couple of years, as physicists and engineers recognized its importance. The most well-studied aspect of this problem is the initial qubit assignment: this is discussed, e.g., in \cite{maslov2008quantum,siraichi2018qubit,van2020mathematical,zhu2020exact}, see also the numerous references therein. The aspect of qubit routing is comparatively less studied, perhaps due to the inherent difficulty of the problem; some notable works in this area are \cite{li2019tackling,mulderij2020polynomial,siraichi2018qubit,van2020mathematical}. Most of the literature proposes heuristic approaches, and exact formulations are limited to certain topologies (e.g., \cite{mulderij2020polynomial}) or to some aspects of the problem (e.g., \cite{bhattacharjee2017depth}). The most significant differences between our work and the existing literature on \emph{exact} solution approaches are therefore: $(i)$ we simultaneously consider the problem of allocating and routing qubits, using a novel network flow formulation with design variables that is unlike any of the previously proposed approaches; $(ii)$ we take into account different costs for inserted SWAPs depending on whether or not they can be merged with adjacent two-qubit gates; $(iii)$ we allow arbitrary hardware topologies. Our paper, like most of the existing literature, considers SWAP-based routing only, which transforms it into a classical problem; in principle, quantum computers can permute qubits without resorting to SWAP gates, and this could yield shorter circuits. In this more general setting, a constant factor improvement can be attained on several graph structures by not limiting routing to SWAPs only \cite{bapat2021quantum}, and this cannot be improved by more than a factor of about 1.5 on the path graph \cite{bapat2020nearly}.

The rest of this paper is organized as follows. Section~\ref{s:preliminaries} sets the context by introducing basic terminology. In Section~\ref{s:formulation} we introduce the proposed mathematical optimization formulation for the qubit allocation problem. In Section~\ref{s:experiments} we present an extensive experimental evaluation of the proposed formulation. Section~\ref{s:conclusions} concludes the paper, summarizing our results.

\section{Preliminaries}
\label{s:preliminaries}
We now introduce the terminology used in this paper, as well as several concepts used in the description of the mathematical optimization problem used in subsequent sections. We distinguish between a {\em logical} quantum circuit, that we want to implement, and a {\em physical} quantum circuit, that is its implementation. A logical quantum circuit is a set of (logical) {\em gates} applied to pairs of (logical) qubits in a certain sequence. It has {\em layers} of gates that can be simultaneously applied (i.e., each logical qubit is involved in at most one gate per layer). Without loss of generality, we consider only two-qubit  gates, assuming other processing steps have expanded multi-qubit gates into single- and two-qubit gates, and absorbing any single-qubit gates into neighboring two-qubit gates. The hardware is described by a graph $H=(V,E)$, where the vertices correspond to physical qubits, and the edges correspond to pairs of qubits onto which a two-qubit gate can be applied. We assume that $H$ has a matching of size larger than or equal to the maximum number of gates in a layer of the logical circuit, so that all gates in a layer can be applied simultaneously. To map the logical circuit to a physical circuit, we perform the following tasks: (i) we create an initial assignment of logical qubits to physical qubits (this is called \emph{qubit assignment}); (ii) we determine a circuit implementing the logical circuit, ensuring that two-qubit gates satisfy the hardware topology by inserting SWAPs if necessary (this is called \emph{qubit routing}); (iii) we determine the final location of the logical qubits in the physical circuit. The combination of these tasks is the \emph{qubit allocation} problem, but the terminology used in the literature is not consistent, and in some papers qubit allocation refers to only some of these tasks. It is not difficult to see that if we allow arbitrary hardware topology and logical circuits, already the qubit assignment problem is a generalization of the $k$-clique problem (i.e., finding a clique of size $k$ or more), therefore it is NP-hard.
 
In this paper we formulate the qubit allocation problem as a binary integer program (BIP). The general form of a BIP is 
$$ 	\min ~ c^{\top} x ~\text{subject to:}~~ Ax = b,~ C x \ge  d,~x\in\{0,1\}^n$$
where  $A\in\R^{m_1\times n},C\in\R^{m_2\times n}, b\in\R^{m_1}, d\in\R^{m_2},$ and $c\in\R^n$. 
The vector $x$ denotes the variables in this formulation, which are only allowed take binary values due to the constraint $x\in\{0,1\}^n$.
The objective is to minimize a linear combination of the variables over the set defined by the equality, inequality and integrality constraints above.

A large number of combinatorial optimization problems can be formulated as a BIP. Indeed, as integer programming is NP-complete, every problem in the complexity class NP can be reduced to it. Numerical algorithms to solve BIPs typically combine solving relaxations of the formulation (the relaxation obtained by dropping the integrality constraint is a linear program, and thus solvable in polynomial time) with enumeration (branch-and-bound).
Even though solving a BIP can take exponential time, because in the worst case we may need to essentially perform enumeration, there has been significant progress in computational methods to solve very large instances of these problems to optimality. For example, the well-known Traveling Salesman Problem can often be solved in a reasonable amount of time even on graphs with many thousands of nodes \cite{tspbook}.
Modern off-the-shelf BIP solvers use techniques such as cutting plane generation, variable bound tightening, symmetry breaking, and primal heuristics  \cite{gurobi,cplex} that are automatically applied to the given input problem instance to produce an optimal solution without searching the (exponential size) solution space explicitly.
In our computational experiments we observe that the combined qubit allocation and qubit routing problem can be solved orders of magnitude faster than explicit enumeration.

\section{Mathematical Formulation}
\label{s:formulation}
	
In this section we formulate the qubit allocation problem as a binary integer program. 
The input to the problem consists of the following: 
\begin{enumerate}
	\item A hardware graph $H=(V,E)$, where  nodes $i\in V$ correspond to physical qubits and  edges $e\in E$, where $e=\{i,j\}$ with $i,j\in V$, correspond to pairs of qubits that can implement two-qubit gates. 
	We do not assume any structure on the hardware graph other than it being connected. 
	\item  A set of logical qubits $Q$ and sequence of gate groups $G=\{G^1,G^2,\ldots, G^m\}$, where each gate group $G^t = \{g^t_{1},g^t_{2},\ldots\}$ is a collection of 2-qubit gates that needs to be implemented at depth (or, time step) $ t\in T=\{1,\ldots,m\}$, and each gate $g^t_{l}=\{p,q\}$ consists of a pair of qubits $p,q\in Q$. 
\end{enumerate}
For simplicity we call logical qubits simply {\em qubits} and physical qubits {\em nodes} from now on.
Without loss of generality, we assume that $|Q|=|V|$ and note that this can be achieved, if necessary, by adding dummy  qubits to $Q$ that do not appear in any gate that needs to be implemented.
We define 	$Q^t=\cup_{k=1}^{|G_t|} g^t_{k} $ to denote the collection of qubits that needs to be implemented at time step $t\in T$. 
Note that we do not assume the gate groups to be non-empty and therefore we allow $G^t = \emptyset$ (and, $Q^t=\emptyset$) for some $t\in T$. In practice, such time steps are inserted to the original  circuit to allow qubit-swaps in order to route qubits in the hardware graph. These  qubit-swaps  are then implemented by adding additional gates to the quantum circuit.
We call such time steps {\em dummy} time steps and denote them with $T^{\text{dummy}}=\{t\in T\::\:G^t =\emptyset\}$.
Finally, for each node $i\in V$, we define a set $N(i)\subseteq V$ to  denote the neighborhood of node $i$, with the interpretation that  qubits can move from node $i$ to any node in $N(i)$ using a single qubit-swap operation.

We start with creating a directed graph $G=(V,A)$ by orienting each edge $e\in E$ of the (undirected) hardware graph $H=(V,E)$ in both directions. More precisely, for each $\{i,j\}\in E$, the set $A$ contains two directed edges $(i,j)$ and $(j,i)$. 
Consequently $|A|=2|E|$. 
In addition, we order the  qubits in each 2-qubit gate in each gate group arbitrarily and fix the ordering.
In this new setting, an (ordered) gate $(p,q)$ can be implemented if $p$ is mapped to a node $i\in V$ and $q$ is mapped to a node $j\in V$ such that $(i,j)\in A$. In this case, the gate can also be implemented if $p$ is mapped to $j$ and $q$ is mapped to $i$ since we have $(j,i)\in A$ as well. 
With this transformation, we can now avoid checking the ``or" condition in (gate $\{p,q\}$ can be implemented on the hardware edge $\{i,j\}$ provided that [$p$ is mapped to a node $i$ and $q$ is mapped to a node $j\in V$] or [$p$ is mapped to $j$ and $q$ is mapped to $i$]) explicitly.

\subsection{Variables and constraints}
\label{s:vars}
We next present the  variables used in our BIP formulation together with the constraints associated with them. The  constraints below define the feasible region of the problem in such a way that there is a one-to-one correspondence between the feasible solutions of the BIP and the solutions to the qubit allocation and routing problem. 
We will discuss several objective functions to be optimized later in  Section~\ref{s:objfun}.
\paragraph{Location variables and constraints:}	 For all $q\in Q$, $i\in V$, and $t\in T$ we define binary variable $w_{q,i}^t \in\{0,1\}$ which takes the value 1 if and only if qubit $q\in Q$ resides at  node $i\in V$ at time $t$. To make sure that each qubit is located at exactly one note at any time step, we have the following constraint:
\begin{equation} \sum_{j\in V} w_{q,j}^t=1\text{~~~for all ~~} q\in Q,~t\in T.	\tag{qubit} \end{equation}
In addition, we have the following constraint to make sure that each node can host exactly one qubit at any time step.
\begin{equation} \sum_{q\in Q} w_{q,j}^t=1\text{~~~for all ~~} j\in V,~t\in T.
\tag{node} \end{equation}
		
\paragraph{Gate variables and constraints:} For all  $t\in T$,  $(p,q)\in G^t$, and $(i,j)\in A$  we define a binary variable $y^t_{(p,q),(i,j)} \in\{0,1\}$ which takes the value 1 if and only if  gate $(p,q)$  is implemented on hardware arc $(i,j)$ at time $t$.
The following constraint makes sure that each gate is implemented exactly once:
\begin{equation} \sum_{(i,j)\in A} y^t_{(p,q),(i,j)} =1  \text{~~~for all ~~} t\in T,~ (p,q)\in G^t.
\tag{gate} \end{equation}
The following (nonlinear) constraint makes sure that a gate can be implemented on the hardware graph if and only if both of its qubits are located at the associated nodes
\begin{equation} y^t_{(p,q),(i,j)} = w_{p,i}^t\times w_{q,j}^t  \text{~~~for all ~~} t\in T,~ (p,q)\in G^t,~ (i,j)\in A. \tag{qubit/gate location} \end{equation}
Each of these constraints can be linearized using the four inequalities below, known as the McCormick inequalities:
\begin{equation} \label{eq:mccormick}\{w_{p,i}^t,~w_{q,j}^t\} ~\ge~ y^t_{(p,q),(i,j)}  ~\ge~ \{0,w_{p,i}^t+ w_{q,j}^t-1\}.\tag{McCormick}\end{equation}
Note that the constraints above imply that $y^t_{(p,q),(i,j)} =1$ if and only if $w_{p,i}^t=1$ and $w_{q,j}^t =1$. 

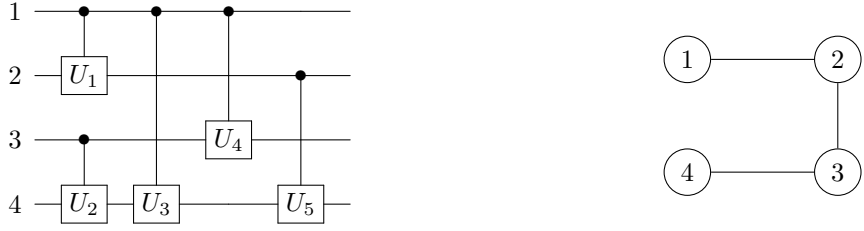
\begin{figure}[t!]
\begin{subfigure}{0.5\textwidth}
\leavevmode
\centering
\Qcircuit @C=1em @R=1em @!R {
\lstick{1} & \ctrl{1} & \ctrl{3} & \ctrl{2} & \qw      & \qw \\
\lstick{2} & \gate{U_1} & \qw      & \qw      & \ctrl{2} & \qw \\
\lstick{3} & \ctrl{1} & \qw      & \gate{U_4} & \qw      & \qw \\
\lstick{4} & \gate{U_2} & \gate{U_3} & \qw      & \gate{U_5} & \qw \\
}
\end{subfigure}
\begin{subfigure}{0.5\textwidth}
    \centering
    \begin{tikzpicture}
    \node[shape=circle,draw=black] (1) at (-.5,0) {1};
    \node[shape=circle,draw=black] (2) at (1.5,0) {2};
    \node[shape=circle,draw=black] (4) at (-.5,-1.5) {4};
    \node[shape=circle,draw=black] (3) at (1.5,-1.5) {3};

    \path [-](1) edge node[above] {} (2);
    \path [-](2) edge node[above] {} (3);
    \path [-](3) edge node[right] {} (4);
    \end{tikzpicture}    
\end{subfigure}
\caption{Quantum circuit and hardware topology used in the example of Figure~\ref{fig:bip_example}.}
\label{fig:circuit_example}
\end{figure}

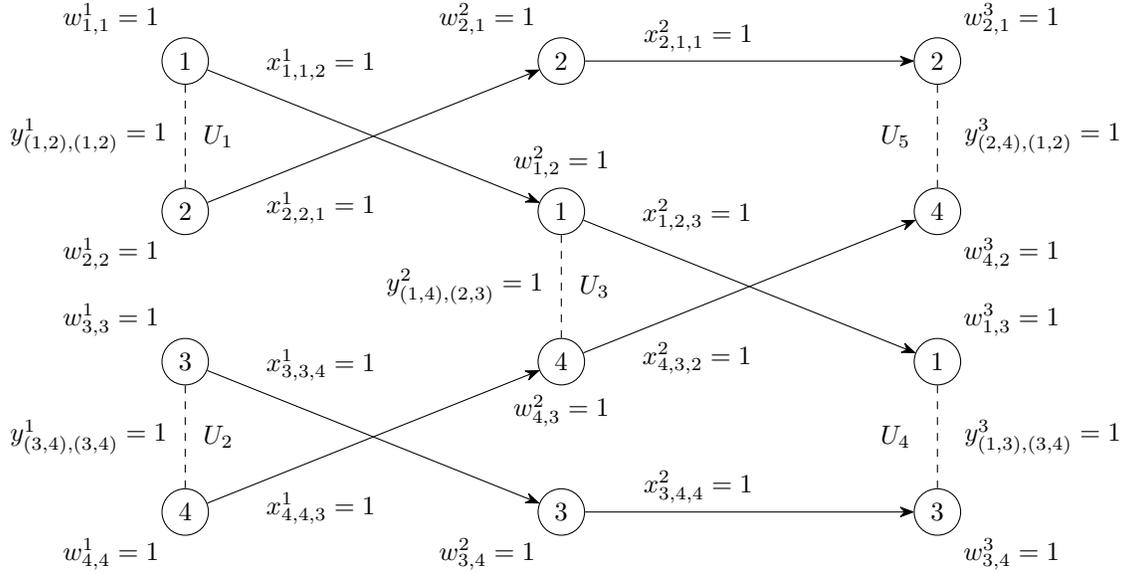
\begin{figure}[tb]
    \centering
    \begin{tikzpicture}
    \node[shape=circle,draw=black,label={above left}:{$w_{1,1}^1=1$}] (1) at (0,0) {1};
    \node[shape=circle,draw=black,label={below left}:{$w_{2,2}^1=1$}] (2) at (0,-2) {2};
    \node[shape=circle,draw=black,label={above left}:{$w_{3,3}^1=1$}] (3) at (0,-4) {3};
    \node[shape=circle,draw=black,label={below left}:{$w_{4,4}^1=1$}] (4) at (0,-6) {4};

    \path [-,dashed](1) edge node[left] {$y_{(1,2),(1,2)}^1 = 1$~~} node[right] {~$U_1$}(2);
    \path [-,dashed](3) edge node[left] {$y_{(3,4),(3,4)}^1 = 1$~~} node[right] {~$U_2$}(4);
    
    \node[shape=circle,draw=black,label={above left}:{$w_{2,1}^2=1$}] (11) at (5,0) {2};
    \node[shape=circle,draw=black,label={above}:{$w_{1,2}^2=1$}] (12) at (5,-2) {1};
    \node[shape=circle,draw=black,label={below}:{$w_{4,3}^2=1$}] (13) at (5,-4) {4};
    \node[shape=circle,draw=black,label={below left}:{$w_{3,4}^2=1$}] (14) at (5,-6) {3};

    \path [-,dashed](12) edge node[left] {$y_{(1,4),(2,3)}^2 = 1$~~} node[right] {~$U_3$}(13);
    
    \path [->,>={Stealth[length=6pt,round]}](1) edge node[pos=0.15,above right] {$x_{1,1,2}^1 = 1$} (12);
    \path [->,>={Stealth[length=6pt,round]}](2) edge node[pos=0.15,below right] {$x_{2,2,1}^1 = 1$} (11);
    \path [->,>={Stealth[length=6pt,round]}](3) edge node[pos=0.15,above right] {$x_{3,3,4}^1 = 1$} (14);
    \path [->,>={Stealth[length=6pt,round]}](4) edge node[pos=0.15,below right] {$x_{4,4,3}^1 = 1$} (13);
    
    \node[shape=circle,draw=black,label={above right}:{$w_{2,1}^3=1$},] (21) at (10,0) {2};
    \node[shape=circle,draw=black,label={below right}:{$w_{4,2}^3=1$}] (22) at (10,-2) {4};
    \node[shape=circle,draw=black,label={above right}:{$w_{1,3}^3=1$}] (23) at (10,-4) {1};
    \node[shape=circle,draw=black,label={below right}:{$w_{3,4}^3=1$}] (24) at (10,-6) {3};

    \path [-,dashed](21) edge node[right] {$~~y_{(2,4),(1,2)}^3 = 1$} node[left] {$U_5$~~~} (22);
    \path [-,dashed](23) edge node[right] {$~~y_{(1,3),(3,4)}^3 = 1$} node[left] {$U_4$~~~} (24);    
        
    \path [->,>={Stealth[length=6pt,round]}](11) edge node[pos=0.15,above right] {$x_{2,1,1}^2 = 1$} (21);
    \path [->,>={Stealth[length=6pt,round]}](12) edge node[pos=0.15,above right] {$x_{1,2,3}^2 = 1$} (23);
    \path [->,>={Stealth[length=6pt,round]}](13) edge node[pos=0.15,below right] {$x_{4,3,2}^2 = 1$} (22);
    \path [->,>={Stealth[length=6pt,round]}](14) edge node[pos=0.15,above right] {$x_{3,4,4}^2 = 1$} (24);
    \end{tikzpicture}    

    \caption{Representation of the decision variables with value 1 of a possible implementation of the circuit in Figure \ref{fig:circuit_example} when executed on a hardware with line topology. The labels inside nodes indicate logical qubits; the physical qubits are numbered 1 to 4 from top to bottom. Note that this implementation is only for illustrative purposes and is not necessarily optimal (e.g., the circuit could be implemented using fewer SWAPs).}
    \label{fig:bip_example}
\end{figure}

\paragraph{Routing variables and constraints:} For all $q\in Q$, $i\in V$, $j\in N(i)\cup\{i\}$, and $t\in T\setminus\{m\}$, we define a binary variable  $x^t_{q,i,j}\in\{0,1\}$ which takes the value 1 if and only if qubit $q\in Q$  is located at node $i$ at time step $t$ and it is  located at node $j$ at the next time step $t+1$. If $i=j$ and $x^t_{q,i,j}=1$ then it means that qubit $q$ does not move and stays at node $i$. 
We model the movement of qubits from one node to another using the following flow balance equations.
\begin{align} w_{q,i}^t&=x^t_{q,i,i} +\sum_{k\in N(i)} x^t_{q,i,k} \text{~~~for all ~~} q\in Q, ~i\in V,~t\in T\setminus\{m\} \tag{out of $w_{q,i}^t$}, \\
w_{q,i}^t&=x^{t-1}_{q,i,i}+\sum_{k\in N(i)} x^{t-1}_{q,k,i} \text{~~~for all ~~} q\in Q, ~i\in V,~t\in T\setminus\{1\}	\tag{in to $w_{q,i}^t$}. \end{align}	
Qubits can move from one node to another under two conditions, depending on whether or not they are involved in a gate at that time step.
If $(p,q)\in G^t$ at time $t\in T$, then qubits $p$ and $q$ can swap positions: \begin{equation}
x^t_{p,i,j} =   x^t_{q,j,i}  \text{~~~for all ~~} t\in T\setminus\{m\},~ (p,q)\in G^t, ~ (i,j)\in A.\end{equation}
Qubits that are not involved in a gate at a time step $t\in T$ can swap positions with a qubit located at one of the neighboring nodes provided that the other qubit is not involved in a gate either:
\begin{equation} \sum_{q\not\in Q^t} x^t_{q,i,j} = \sum_{p\not\in Q^t} x^t_{p,j,i}
\text{~~~for all ~~} t\in T\setminus\{m\},~  (i,j)\in A.\end{equation}
Using the definition of $w_{q,i}^t$ in terms of the routing variables, for all $t \in T \setminus \{m\}$ we can strengthen the inequalities \eqref{eq:mccormick}. In their place, we use a version that exploits knowledge of the fact that qubits involved in a gate can only swap with each other:
\begin{equation} \label{eq:mccormickstr}\{x_{p,i,i}^t+x_{p,i,j}^t,~x_{q,j,i}^t+x_{q,j,j}^t\} ~\ge~ y^t_{(p,q),(i,j)}  ~\ge~ \{0,w_{p,i}^t+ w_{q,j}^t-1\}.\tag{McCormickStr}\end{equation}
These inequalities cannot be used at time $t = m$ because the corresponding routing variables $x$ are not defined. We use \eqref{eq:mccormickstr} in the computational experiments.
An illustration of the meaning of the decision variables for a small circuit is given in Figures~\ref{fig:circuit_example}-\ref{fig:bip_example}.

\paragraph{Counting dummy time steps:}
In some cases, not all dummy time steps are needed to implement the quantum circuit. For all $t\in T^{\text{dummy}}$ we define a binary variable  $z^t\in\{0,1\}$ which takes the value 1 if some qubit-swap takes place at that time step and write the following constraint:
\begin{equation}\label{eq:dummytimestep}\sum_{(i,j)\in A}x^t_{q,i,j}\le z^t\qquad \forall {q\in Q}, ~\forall t\in T^{\text{dummy}}  \end{equation}
These variables and the associated constraints are not needed to model the problem correctly but they are used in some of the cost functions that we introduce later. Notice that if $z^t =0$ for some $t\in T$, then for all $q$ and $i$ we have $x^t_{q,i,i} = 1$ and therefore none of the logical qubits move at time step $t$.

Now consider two time steps $t_1,t_2\in T$ such that $t_1<t_2-2$  and $t\in T^{\text{dummy}}$ for all $t_1<t<t_2$. In this  case we can impose the following constraints without loss of generality:
\begin{equation} z^{t_1+1}~\ge~ z^{t_1+2}~\ge\cdots\ge~ z^{t_2-1}. \end{equation}
These constraints are not needed to model the problem correctly but they help break {\em symmetry} in the integer program and lead to better computational performance.


\subsection{Objective functions}
\label{s:objfun}
The primary goal of solving the qubit assignment problem is to find a circuit implementation that maximizes the circuit fidelity while minimizing the use of quantum resources. These two objectives can be mathematically characterized in multiple ways. On top of the difficulty of choosing the right mathematical definition, in the context of this paper we are limited by the fact that we ultimately want to end up with objective functions that can be efficiently optimized with existing mixed-integer programming software. Taking this aspect into consideration, we consider three possible objective functions; here we described their rationale, and how they can be formulated using the mathematical optimization model described in the previous section.

	\paragraph{Minimizing error rate:} The first objective function that we consider aims to maximize the probability of successful execution of the quantum circuit under simplified assumptions.
	To this end, we assume that: (i) a quantum circuit is executed successfully if all of its gates execute successfully, and failures of the gates are independent events; (ii) the success probability of a gate on a given edge only depends on the gate itself, the edge, and the number of CNOT gates used to implement the gate. 
	This is a simplified model that does not take into account any temporal or spatial correlations. Under this simplified model we obtain a proxy for the circuit fidelity. The proxy has the property that, after a log transformation, it yields a linear function that can be maximized within our BIP model. Even if the simplifying assumptions are not satisfied in practice, we expect the chosen proxy to correlate well with some measure of fidelity of the circuit (this is experimentally investigated in Section~\ref{s:experiments}).

	With these assumptions in mind,  let $\beta_{ij}$ denote the fidelity of a CNOT gate applied on edge $(i,j)\in A$, and let ${\cal F}(g, k)$ denote the fidelity of implementing gate $g\in G$ using at most $k$ CNOT gates. 
	As each CNOT gate succeeds with probability $\beta_{ij}$, and the implementation of gate $g$ with $k$ CNOTs succeeds with probability ${\cal F}(g, k)$, the  probability of success for the gate when $k$ CNOTs are employed becomes ${\cal F}(g, k) \beta_{ij}^k$.
	Therefore, as a preprocessing step, we first determine how many CNOTs should be used for gate $g$ if it is  the implemented on edge $(i,j)\in A$:	
	\begin{equation*}
		n(g, (i,j)) := \arg \max_{k=0,1,2,3} \big\{ {\cal F}(g, k) \beta_{ij}^k \big\},
	\end{equation*}
and define the best success probability of implementing gate $g\in G$  on edge $(i,j)\in A$ as follows:
	\begin{equation*}
		{\cal P}^*(g, (i,j)) :=  \max_{k=0,1,2,3} \big\{ {\cal F}(g, k) \beta_{ij}^k \big\}
		=\left( {\cal F}(g, n(g, (i,j))) \beta_{ij}^{n(g, (i,j))} \right). 
	\end{equation*}

	Similarly, for each gate $g\in G$, we define  $g_{\text{SWAP}}(g, k)$ to denote the fidelity of implementing gate $g\in G$ using at most $k$ CNOT gates when the gate is followed by a SWAP on the same pair of qubits. We can then determine the optimal number of  CNOTs  and define the best success probability of implementing gate $g\in G$  on edge $(i,j)\in A$ with a SWAP as follows:
	\begin{equation*}
		{\cal P}_{\text{SWAP}}^*(g, (i,j))  :=  \max_{k=0,1,2,3} \big\{ {\cal F}(g_{\text{SWAP}}, k) \beta_{ij}^k \big\}.
    \end{equation*}

	Therefore, the logarithm of the  probability of success of the circuit can now be expressed as:

	\begin{align}
		&\sum_{t \in T} \sum_{ g = (p, q) \in G^t} \sum_{(i,j) \in A}  \log \left( {\cal P}^*(g, (i,j))\right) \left[y^t_{(p,q),(i,j)} - \frac{x^t_{p,i,j} + x^t_{q,j,i}}{2} \right]\qquad\qquad \notag\\
		&\qquad+~\sum_{t \in T} \sum_{ g = (p, q) \in G^t} \sum_{(i,j) \in A} 
		\log \left( {\cal P}_{\text{SWAP}}^*(g, (i,j))  \right) \left[\frac{x^t_{p,i,j} 
	+ x^t_{q,j,i}}{2}\right]\qquad\qquad \label{eq:objerrorrate}\\
		&\qquad\qquad+~\	\sum_{t \in T}  \sum_{q \in Q \setminus Q^t} \sum_{(i,j) \in A}  \log \left( \beta_{ij}^{3/2}  \right) x^t_{q,i,j}\qquad\qquad\notag
	\end{align}
where $Q^t \subseteq Q$ denotes the set of logical qubits involved in some gate at time step $t\in T$, i.e., $$Q^t := \{q \in Q : \exists (p,q) \in G^t \text{ or } \exists (q,p) \in G^t\}.$$

	To see that this expression \eqref{eq:objerrorrate} above correctly models the logarithm of the probability of success, 
	notice that for time step $t\in T$  the variable $y^t_{(p,q),(i,j)}$ in our model takes value 1 if and only if gate $g = (p,q)\in G^t$ is implemented on arc $(i,j)\in A$ (i.e. qubit $p$ is in node $i$ and qubit $q$ is in node $j$). In this case, 
	\begin{itemize}		
	\item	If  qubits $p$ and $q$ do not swap positions then $x^t_{p,i,j} = x^t_{q,j,i} = 0$ and the probability of success of this gate is precisely ${\cal P}^*(g, (i,j)) = {\cal F}(g, n(g, (i,j))) \beta_{ij}^{n(g, (i,j))}$.
	\item If, on the other hand, these qubits swap positions, and $x^t_{p,i,j} = x^t_{q,j,i} = 1$. The probability of success for this gate is ${\cal P}_{\text{SWAP}}^*(g, (i,j)) = {\cal F}(g_{\text{SWAP}}, n(g_{\text{SWAP}}, (i,j))) \beta_{ij}^{n(g_{\text{SWAP}}, (i,j))}$.
		\	\end{itemize}

For all qubits $q\in Q\setminus Q^t$ that are not involved in a gate at time step $t$, if $x^t_{q,i,j} = 1$ then the qubits at nodes $i$ and $j$ swap positions; this SWAP has probability of success $\beta_{ij}^3$ (note that we use the coefficient $\beta_{ij}^{3/2}$ in the expression of the objective function, because if $x^t_{q,i,j}=1$ also $x^t_{q',j,i} = 1$ for some $q'$, and each SWAP should be counted once).

Under our simplifying assumption, the circuit is successful if all its gates are successful, hence the total probability of success is the product of all the associated success probabilities. 
We transform this product into a linear expression by taking the logarithm,  yielding the linear function in \eqref{eq:objerrorrate}, which we maximize as the objective function of the BIP. Since we express all objective functions as minimization problems, for consistency and ease of interpreting the plots obtained from experimental results, we transform the maximization of the success probability into the (equivalent) minimization of the error rate.

\paragraph{Minimizing circuit depth:} The second objective function that we consider is the depth of the circuit, which we use as a proxy for its execution time, as deeper circuits are more likely to be affected by noise and errors from different sources. Thus, we aim to minimize circuit depth. We can assume without loss of generality that the circuit $G$ requires at least $|\{ t : G^t \neq \emptyset, t \in T\}|$ layers of gates to be implemented, because the logical circuit can be preprocessed to ensure all layers are strictly necessary (using the assumption that at any given layer, each qubit can only be involved in one gate). Hence, it is sufficient to minimize the number of dummy steps. Using constraints \eqref{eq:dummytimestep}, this can be formulated as:
\begin{equation*}
\label{eq:objdepth}
\min  \sum_{t \in T^{\text{dummy}}} z^t.
\end{equation*}

\paragraph{Minimizing crosstalk:} The third objective function that we consider is the minimization of crosstalk, i.e., interference due to simultaneous use of certain edge pairs. Crosstalk is a major source of noise in existing devices \cite{murali2020software}. Characterizing hardware crosstalk is a difficult task, but it is outside the scope of this paper. Here we assume that crosstalk on the device has already been characterized, and we are given a set $X$ of edge pairs that exhibit crosstalk. The goal is to implement the circuit minimizing the number of layers at which edges that form a pair in $X$ are simultaneously used. In other words, we look at the amount of crosstalking edge pairs that are used in the course of the circuit: for $t \in T$ and for $(e_1, e_2) \in X$, if edges $e_1$ and $e_2$ are both employed at time step $t$, the amount of crosstalking edge pairs is increased by one. We can minimize this amount as follows:
\begin{equation}
\label{eq:objcrosstalk}
    \min \sum_{t \in T} \sum_{(e_1, e_2) \in X} u^t_{e_1} u^t_{e_2},
\end{equation}
where for $t \in T, e \in E$, $u^t_e = 1$ if and only if edge $e$ is used at time step $t$. Note that we must ensure that the variables $u^t_e$ are correctly defined in terms of the other variables via constraints; this is ensured by adding the constraints for all $ t\in T,~ e=\{i,j\}\in E$:
\begin{equation*}
    u^t_{e} = \sum_{(p,q) \in G^t} y^t_{(p,q),(i,j)} + \sum_{p \in Q \setminus Q^t} x^t_{q,i,j} +\sum_{(p,q) \in G^t} y^t_{(p,q),(j,i)} + \sum_{p \in Q \setminus Q^t} x^t_{q,j,i}. 
\end{equation*}
Furthermore, note that this objective function is quadratic. Commercial integer programming solvers typically handle quadratic objective functions, so in principle we do not need to take any steps. However, in practice we found that the numerical performance of the solver increases if each product of binary variables is linearized, in the same way as indicated in \eqref{eq:mccormick}. This linearization does not have to be carried out explicitly when using IBM ILOG CPLEX, as it is carried out automatically by setting the appropriate option at runtime; we follow this approach. We remark that objective function \eqref{eq:objcrosstalk} is only one of multiple ways to model the minimization of crosstalk. In particular, the proposed formulation has the advantage of simplicity: it does not require quantifying the strength of the crosstalk interactions; rather, we are only required to indicate which pairs of edges should be excluded from simultaneous usage if possible. Of course, if a more accurate characterization of crosstalk, that includes quantitative information, is available, then alternative approaches should be considered.

\section{Experimental evaluation}
\label{s:experiments}
In this section we  present an extensive experimental evaluation of our methodology  to solve the qubit allocation problem. All optimization problems are solved with IBM ILOG CPLEX 12.10 \cite{cplex} on 8 cores of a virtual Intel Xeon Platinum 8260 CPUs clocked at 2.40GHz, instantiated on an IBM cloud. 

We use Qiskit \cite{qiskit} to create the circuits and a custom transpiler that uses our own code to solve the qubit assignment and routing problems. To assess the fidelity of an implementation $U$ of a target unitary $U_{\text{target}}$ (i.e., the function ${\cal F}$ used in Section~\ref{s:objfun}) we use a formula derived from \cite{horodecki1999general}:
\begin{equation*}
    \frac{d + \left|\text{Tr}(U_\text{target} U^\dag)\right|^2}{d(d+1)},
\end{equation*}
where $d$ is the dimension of the unitary, i.e., $d=4$ in our case. We fix the fidelity of a CNOT gate $\beta_{ij}$ over edge $(i,j)$ to 0.9936, which is the average CNOT fidelity on the devices used in \cite{jurcevic2021demonstration}, and is therefore a realistic estimate. (As detailed in Section~\ref{s:objfun}, our approach supports assigning a different CNOT fidelity to each edge, if such data is available.)

\subsection{Quantum volume circuits}
\label{s:qv}
Quantum volume (QV) is a holistic, single-number metric used to benchmark near-term quantum computers of modest size ($n\lesssim 50$) \cite{cross2019validating}. The QV protocol yields the answer to the question `what is the largest, random square circuit that can be successfully implemented on noisy hardware'. A square circuit is a circuit of equal width $w$ and depth $d$, where width is the number of qubits in the quantum register and depth is the number of layers. Each layer consists of a random permutation of the qubit register followed by $\lfloor w/2 \rfloor$ random $SU(4)$'s over the Haar measure. The random permutations ensure that the two-qubit gate operations are applied on random pairs in each layer of the circuits, hence highlighting the importance of routing and layout algorithms for sparsely connected hardware implementations. For each circuit we compute the ideal probability distribution and determine the heavy output as the set of bit-strings with their corresponding probability being above the median probability value. Each circuit is then executed at least once on real hardware or on a noisy simulator and if the recorded outcome bit-string is heavy, i.e. is within the set of the pre-computed ideal heavy output bit-strings, we mark this run as successful. In order to pass the quantum volume test of size $2^w=2^d$, one has to execute $ >100$ random circuits with the heavy output probability (HOP) $ > 2/3$ within $2\sigma$ of the one-sided binomial confidence interval \cite{cross2019validating}. Here, we focus on a pre-drawn set of random circuits and compare them with different routing/layout routines against each other, hence eliminating the uncertainty due to the randomness of generating circuits. We execute each circuit $n_s$ times and estimate the heavy output probability for each individual circuit. Finally, we compute the average heavy output probability taking the mean of all individual heavy outputs, and use it as a benchmark to compare the effects of the different qubit allocation algorithms on actual hardware. From now on, we use ``HOP'' of a set of circuits to denote the average heavy output probability obtained as the mean of individual heavy outputs.

\subsection{Hardware topology}

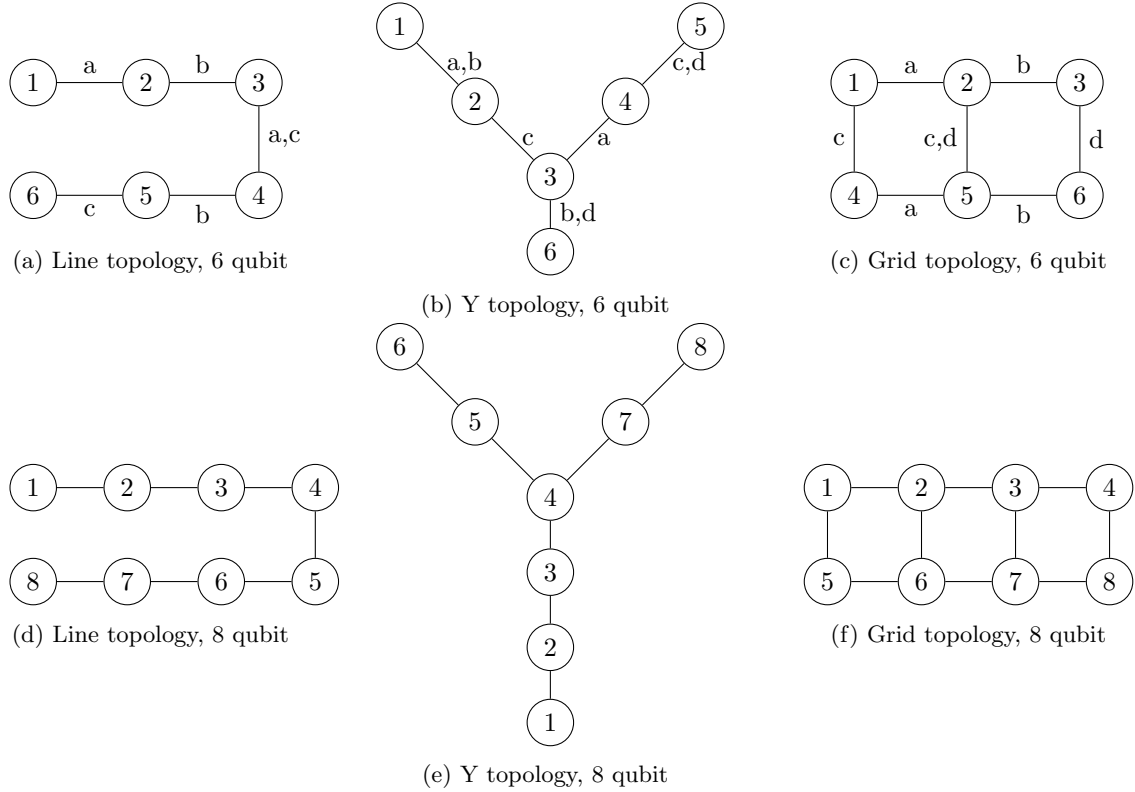
\begin{figure}[tb]
    \centering
    \begin{subfigure}{0.25\textwidth}
    \centering
    \begin{tikzpicture}
    \node[shape=circle,draw=black] (1) at (0,0) {1};
    \node[shape=circle,draw=black] (2) at (1.5,0) {2};
    \node[shape=circle,draw=black] (3) at (3,0) {3};
    \node[shape=circle,draw=black] (6) at (0,-1.5) {6};
    \node[shape=circle,draw=black] (5) at (1.5,-1.5) {5};
    \node[shape=circle,draw=black] (4) at (3,-1.5) {4};

    \path [-](1) edge node[above] {a} (2);
    \path [-](2) edge node[above] {b} (3);
    \path [-](3) edge node[right] {a,c} (4);
    \path [-](4) edge node[below] {b} (5);
    \path [-](5) edge node[below] {c} (6);
    \end{tikzpicture}    
    \caption{Line topology, 6 qubit}
    \label{fig:6q_line}
    \end{subfigure}
    \hfill
    \begin{subfigure}{0.3\textwidth}
    \centering
    \begin{tikzpicture}
    \node[shape=circle,draw=black] (1) at (-2,0) {1};
    \node[shape=circle,draw=black] (2) at (-1,-1) {2};
    \node[shape=circle,draw=black] (3) at (0,-2) {3};
    \node[shape=circle,draw=black] (4) at (1,-1) {4};
    \node[shape=circle,draw=black] (5) at (2,0) {5};
    \node[shape=circle,draw=black] (6) at (0,-3) {6};

    \path [-](1) edge node[above,right] {a,b} (2);
    \path [-](2) edge node[above,right] {c} (3);
    \path [-](3) edge node[below,right] {a} (4);
    \path [-](4) edge node[below,right] {c,d} (5);
    \path [-](3) edge node[right] {b,d} (6);
    \end{tikzpicture}    
    \caption{Y topology, 6 qubit}
    \label{fig:6q_y}
    \end{subfigure}
    \hfill
    \begin{subfigure}{0.3\textwidth}
    \centering
    \begin{tikzpicture}
    \node[shape=circle,draw=black] (1) at (0,0) {1};
    \node[shape=circle,draw=black] (2) at (1.5,0) {2};
    \node[shape=circle,draw=black] (3) at (3,0) {3};
    \node[shape=circle,draw=black] (4) at (0,-1.5) {4};
    \node[shape=circle,draw=black] (5) at (1.5,-1.5) {5};
    \node[shape=circle,draw=black] (6) at (3,-1.5) {6};

    \path [-](1) edge node[above] {a} (2);
    \path [-](2) edge node[above] {b} (3);
    \path [-](1) edge node[left] {c} (4);
    \path [-](2) edge node[left] {c,d} (5);
    \path [-](3) edge node[right] {d} (6);
    \path [-](4) edge node[below] {a} (5);
    \path [-](5) edge node[below] {b} (6);
    \end{tikzpicture}    
    \caption{Grid topology, 6 qubit}
    \label{fig:6q_grid}
    \end{subfigure}
    
        \begin{subfigure}{0.25\textwidth}
    \centering
    \begin{tikzpicture}
    \node[shape=circle,draw=black] (1) at (0,0) {1};
    \node[shape=circle,draw=black] (2) at (1.25,0) {2};
    \node[shape=circle,draw=black] (3) at (2.5,0) {3};
    \node[shape=circle,draw=black] (4) at (3.75,0) {4};
    \node[shape=circle,draw=black] (8) at (0,-1.25) {8};
    \node[shape=circle,draw=black] (7) at (1.25,-1.25) {7};
    \node[shape=circle,draw=black] (6) at (2.5,-1.25) {6};
    \node[shape=circle,draw=black] (5) at (3.75,-1.25) {5};

    \path [-](1) edge node[above] {} (2);
    \path [-](2) edge node[above] {} (3);
    \path [-](3) edge node[right] {} (4);
    \path [-](4) edge node[below] {} (5);
    \path [-](5) edge node[below] {} (6);
    \path [-](6) edge node[below] {} (7);
    \path [-](7) edge node[below] {} (8);
    \end{tikzpicture}    
    \caption{Line topology, 8 qubit}
    \label{fig:8q_line}
    \end{subfigure}
    \hfill
    \begin{subfigure}{0.3\textwidth}
    \centering
    \begin{tikzpicture}
    \node[shape=circle,draw=black] (6) at (-2,0) {6};
    \node[shape=circle,draw=black] (5) at (-1,-1) {5};
    \node[shape=circle,draw=black] (4) at (0,-2) {4};
    \node[shape=circle,draw=black] (7) at (1,-1) {7};
    \node[shape=circle,draw=black] (8) at (2,0) {8};
    \node[shape=circle,draw=black] (3) at (0,-3) {3};
    \node[shape=circle,draw=black] (2) at (0,-4) {2};
    \node[shape=circle,draw=black] (1) at (0,-5) {1};

    \path [-](1) edge node[above,right] {} (2);
    \path [-](2) edge node[above,right] {} (3);
    \path [-](3) edge node[below,right] {} (4);
    \path [-](4) edge node[below,right] {} (5);
    \path [-](5) edge node[below,right] {} (6);
    \path [-](4) edge node[below,right] {} (7);
    \path [-](7) edge node[below,right] {} (8);
    \end{tikzpicture}    
    \caption{Y topology, 8 qubit}
    \label{fig:8q_y}
    \end{subfigure}
    \hfill
    \begin{subfigure}{0.3\textwidth}
    \centering
    \begin{tikzpicture}
    \node[shape=circle,draw=black] (1) at (0,0) {1};
    \node[shape=circle,draw=black] (2) at (1.25,0) {2};
    \node[shape=circle,draw=black] (3) at (2.5,0) {3};
    \node[shape=circle,draw=black] (4) at (3.75,0) {4};
    \node[shape=circle,draw=black] (5) at (0,-1.25) {5};
    \node[shape=circle,draw=black] (6) at (1.25,-1.25) {6};
    \node[shape=circle,draw=black] (7) at (2.5,-1.25) {7};
    \node[shape=circle,draw=black] (8) at (3.75,-1.25) {8};
    
    \path [-](1) edge node[above] {} (2);
    \path [-](2) edge node[above] {} (3);
    \path [-](3) edge node[above] {} (4);
    \path [-](1) edge node[left] {} (5);
    \path [-](2) edge node[left] {} (6);
    \path [-](3) edge node[right] {} (7);
    \path [-](4) edge node[right] {} (8);
    \path [-](5) edge node[below] {} (6);
    \path [-](6) edge node[below] {} (7);
    \path [-](7) edge node[below] {} (8);
    \end{tikzpicture}    
    \caption{Grid topology, 8 qubit}
    \label{fig:8q_grid}
    \end{subfigure}
    \caption{Hardware topologies used in the experimental evaluation. In Section~\ref{s:pareto} we assume that the pairs of cross-talking edges are as indicated by the alphabetic labels: each label appears on exactly two edges in each graph, and those two edges represent a cross-talking pair.}
    \label{fig:topologies}
\end{figure}

To study the behavior of the algorithms for the qubit allocation problem under different conditions, we consider $6$- and $8$-qubit hardware, each with three different topologies: a chain, a grid, and a Y-shaped graph. All hardware graphs are depicted in Figure \ref{fig:topologies}. The chain topology is chosen because it is widely available on IBM chips; the grid, on the other hand, is chosen because it is one of the densest topologies that can be conceivably constructed in near-term devices, therefore it allows us to test our algorithms under very different conditions. In particular, we expect that mapping a QV circuit onto a grid is considerably easier than mapping it onto a chain, while the Y topology is somewhere in between, allowing us to study the difficulty of optimizing the three considered objective functions in different settings. For the 6-qubit topologies in Figure \ref{fig:topologies}, we also indicate which pairs of edges are assumed to exhibit cross-talk effects for the purpose of some experiments.

\subsection{Choice of objective function for the integer program}
\label{s:pareto}
We discuss three possible objective functions in Section~\ref{s:objfun}. Ideally, we would like to optimize all of them simultaneously as multiobjective optimization problem; however, generating the set of all Pareto-optimal solutions with three objectives is an arduous task, and in practical situations one would like to have a simple strategy that is not too burdensome. We therefore design a preliminary experiment to study the trade-offs between the three objective functions, and use the results of this experiment to distill some guidelines that inform our subsequent experiments.

In this experiment, we look at the effect of optimizing the three objective functions discussed in Section~\ref{s:objfun} sequentially, in different orders. The methodology is the following. We consider the six possible orders of three objective functions. Then, for a given circuit and a given order of the objective functions, say, $O_1(x), O_2(x), O_3(x)$, which we assume to be minimization problems for simplicity of illustration:
\begin{enumerate}
    \item We optimize $O_1(x)$, and record its optimal value $o_1$;
    \item We add the constraint $O_1(x) \le o_1$ to the formulation, optimize $O_2(x)$, and record its optimal value $o_2$;
    \item We add the constraint $O_2(x) \le o_2$ to the formulation, optimize $O_3(x)$, and record its optimal value $o_3$.
\end{enumerate}
This is equivalent to requiring that each solution belongs to the optimal face\footnote{The feasible region of a mixed-integer linear program is a polyhedron, hence any halfspace $a^{\top} x \le b$ that touches the boundary defines a face.} of the feasible region for a given objective function, then sequentially optimizing the remaining objective functions. We repeat steps 2 and 3 above several times, increasing the value of $o_1$ at the r.h.s.~of the constraint $O_1(x) \le o_1$, so as to generate an approximation of the Pareto front for different values of the first objective.

For each topology we run 200 randomly generated 6-qubit quantum volume circuits. We limit our attention to 6-qubit quantum volume circuits because we have to solve several thousand instances of the qubit allocation problem, hence we need to keep the computation times under control: for 6-qubit quantum volume circuits, each instance is typically solved to proven global optimality in less than one minute. The number of dummy time steps introduced between every two original circuit layers is set to 5; this ensures that every qubit permutation is reachable in between layers.

\begin{figure}
    \centering
    \begin{subfigure}{0.5\textwidth}
    \centering
    \includegraphics[width=.9\linewidth]{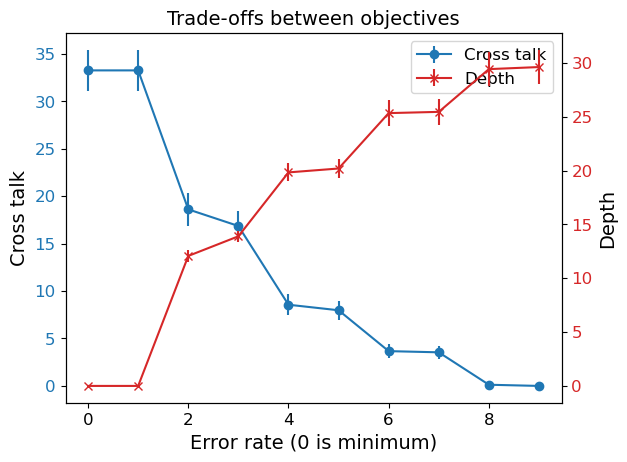}
    \caption{Error rate vs.\ cross talk and depth}
    \label{fig:pareto_ec_y_selected}
    \end{subfigure}%
    \begin{subfigure}{0.5\textwidth}
    \centering
    \includegraphics[width=.9\linewidth]{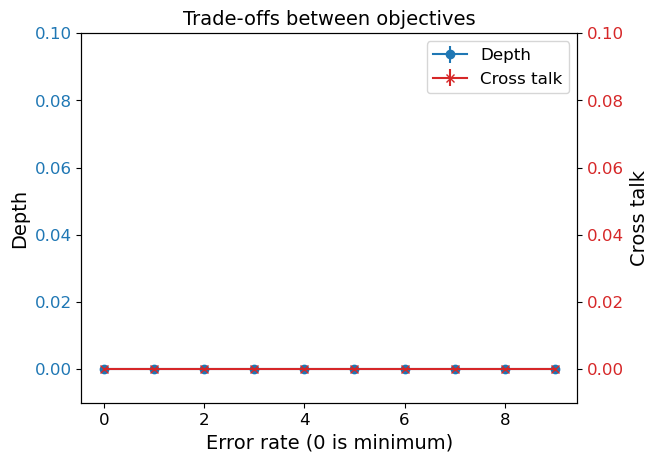}
    \caption{Error rate vs.\ depth and cross talk}
    \label{fig:pareto_ed_y_selected}
    \end{subfigure}
    \caption{Pareto curves when optimizing three objectives in two different orders, on a Y-shaped topology. The $x$ axis contains the first objective to be optimized, with 0 indicating the global optimum. The left $y$ axis contains the second objective to be optimized. The right $y$ axis contains the third objective to be optimized. All $y$ values reported are relative increase with respect to the minimum. Vertical bars denote $\pm$ standard error.}
    \label{fig:pareto_selected}
\end{figure}

Results are reported in Figures~\ref{fig:pareto_line}, \ref{fig:pareto_y} and \ref{fig:pareto_grid} in the Appendix. Each figure contains 6 subfigures corresponding to the six possible orders of the three objective functions. The objective function on the $x$ axis is the first to be optimized, then its value is increased (and set as a constraint, as described above) to observe the effect on the other objective functions. Depth and cross-talk are increased in steps of 1; the error rate objective function is increased in steps equivalent to the cost of a single SWAP gate on the device. Note that we use different scales on the left and right $y$ axis, because each reports a different objective function. For illustrative purposes, we report two subplots of Figure \ref{fig:pareto_y} in the main text, in Figure~\ref{fig:pareto_selected}.

These plots lead to the following observations. The most important fact is that error rate and depth can be simultaneously minimized on the three tested topologies: when error rate and depth are minimized as the first two objectives, the corresponding lines in the plots are flat, indicating that both attain their global minimum value, and there is no trade-off between the two objectives to be paid. On the other hand, cross-talk imposes some trade-offs with the other two objectives. This is evident, for example, in Figure~\ref{fig:pareto_ec_y_selected}. In these circuits one may have to accept up to 4 uses of pairs of edges that interfere with each other, to obtain a circuit with globally minimum depth and error rate objective. Our observations indicate that the price for avoiding certain edge pairs can be significant in terms of depth and required SWAPs. We remark that our tests assume that there are several ``forbidden'' edge pairs in each topology, and the situation could be remarkably different if the number of such edge pairs decreases.

While we do not expect that the simultaneous minimization of error rate and depth (without trade-offs) can be carried out for the three tested topologies regardless of the size of the circuit, we empirically observed that this is the case for 8-qubit quantum volume circuits, in addition to the 6-qubit quantum volume circuits discussed above. Thus, at least for small circuit sizes, these two objectives can be pursued at the same time; this implies that optimizing a linear combination of the two objectives with appropriately chosen weights would also yield a solution that globally minimizes both. In the rest of the paper, we minimize the error rate as the first objective, fix its optimal value, and minimize depth next. This is the default strategy in our implementation, for the reason outlined above and also because, as mentioned, assessing cross talk is a complicated process and is not routinely performed on the IBM devices used in our experimentation.

\subsection{Comparison of solution algorithms}
\label{s:comparison}
We test and compare five different approaches to solve the qubit allocation problem:
\begin{itemize}
    \item The BIP formulation (label ``BIP''), minimizing error rate as the first objective, and depth as the second objective.
    \item Qiskit's default qubit allocation algorithm SABRE (label ``SABRE'').
    \item A hybrid algorithm that uses BIP to obtain the initial qubit assignment, then calls SABRE to solve the qubit routing problem (label ``BIP-Layout'').
    \item A hybrid algorithm that uses SABRE to obtain the initial qubit assignment, then calls CIP to solve the qubit routing problem (label ``BIP-Routing'').
    \item The BIP formulation with the additional constraint that the final qubit permutation must be the same as the initial qubit assignment (label ``BIP-Constrained'').
\end{itemize}
We test BIP-Constrained because SABRE has a similar restriction\footnote{To the best of our knowledge, this is a design decision motivated by the fact that implementing algorithms that employ circuits as black-box oracles is generally easier, if one can assume that qubits are not permuted by the black box.}, and we want to assess the impact of this restriction on the solution of the qubit allocation problem.

These five algorithms are tested on the same three hardware topologies discussed above: line, Y, and grid. We discuss results for 6-qubit volume circuits and 8-qubit volume circuits separately.

\subsubsection{6-qubit volume circuits}
\begin{figure}[tp]
    \centering
    \begin{subfigure}{0.86\textwidth}
    \centering
    \includegraphics[width=\linewidth]{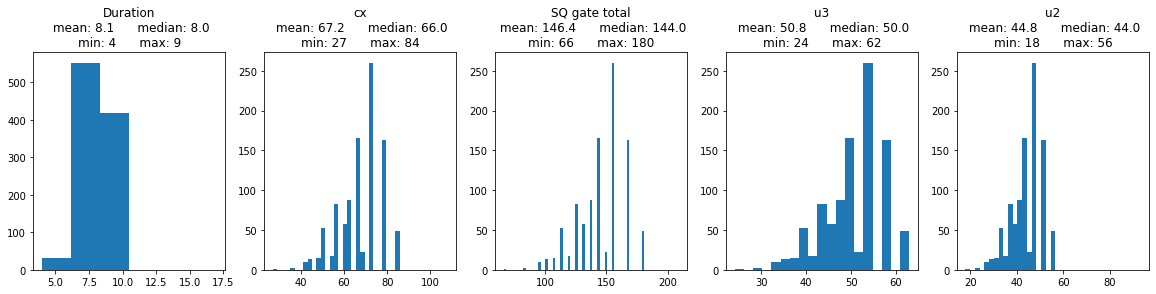}
    \caption{BIP}
    \label{fig:6q_compiled_line_bip}
    \end{subfigure}\\
    \begin{subfigure}{0.86\textwidth}
    \centering
    \includegraphics[width=\linewidth]{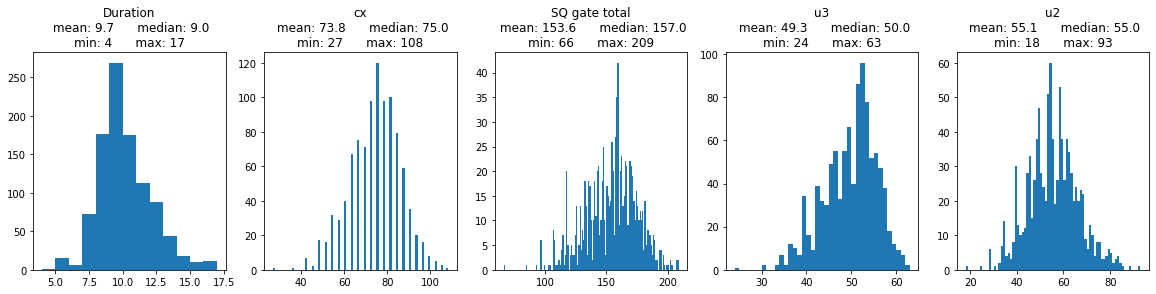}
    \caption{SABRE}
    \label{fig:6q_compiled_line_sabre}
    \end{subfigure}\\
    \begin{subfigure}{0.86\textwidth}
    \centering
    \includegraphics[width=\linewidth]{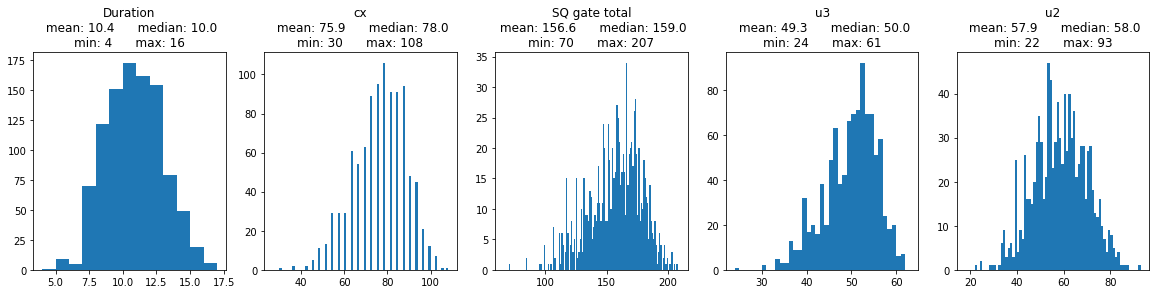}
    \caption{BIP-Layout}
    \label{fig:6q_compiled_line_bip_layout}
    \end{subfigure}\\
    \begin{subfigure}{0.86\textwidth}
    \centering
    \includegraphics[width=\linewidth]{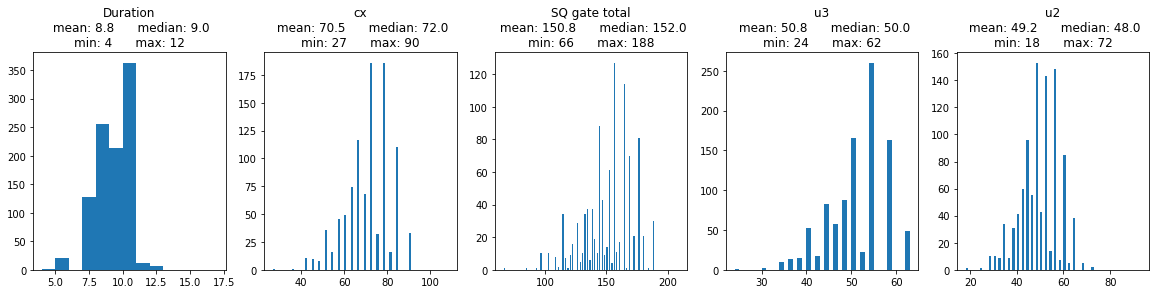}
    \caption{BIP-Routing}
    \label{fig:6q_compiled_line_bip_routing}
    \end{subfigure}\\
    \begin{subfigure}{0.86\textwidth}
    \centering
    \includegraphics[width=\linewidth]{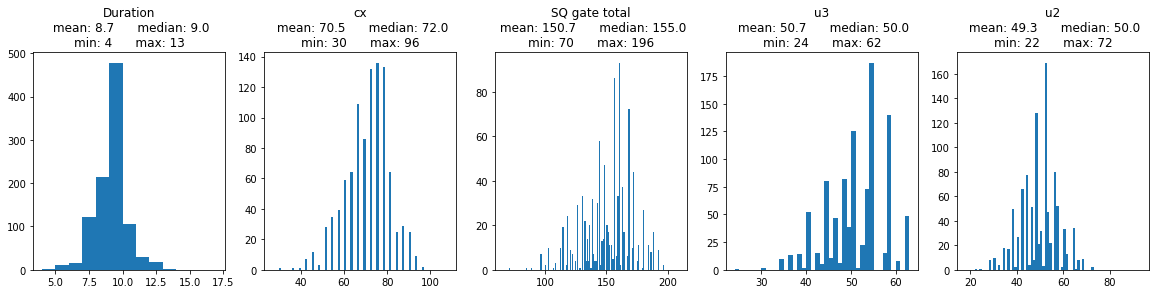}
    \caption{BIP-Constrained}
    \label{fig:6q_compiled_line_bip_constrained}
    \end{subfigure}
    \caption{Statistics for 6-qubit volume circuits compiled on the line topology using different algorithms to solve the qubit allocation problem.}
    \label{fig:6q_compiled_line}
\end{figure}

\begin{figure}[tp]
    \centering
    \begin{subfigure}{0.86\textwidth}
    \centering
    \includegraphics[width=\linewidth]{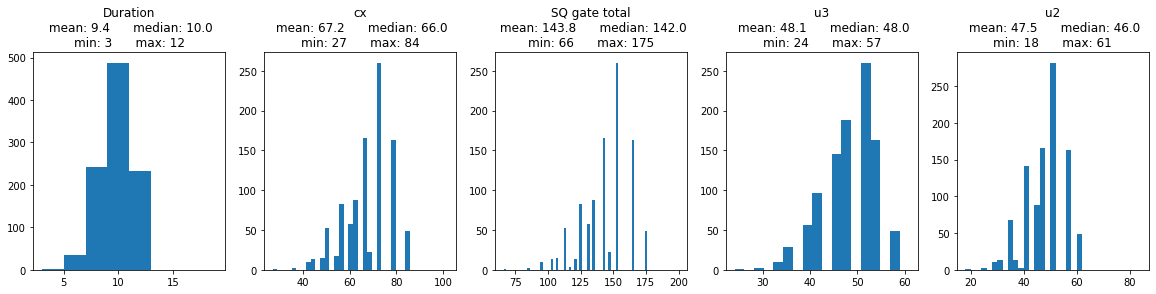}
    \caption{BIP}
    \label{fig:6q_compiled_y_bip}
    \end{subfigure}\\
    \begin{subfigure}{0.86\textwidth}
    \centering
    \includegraphics[width=\linewidth]{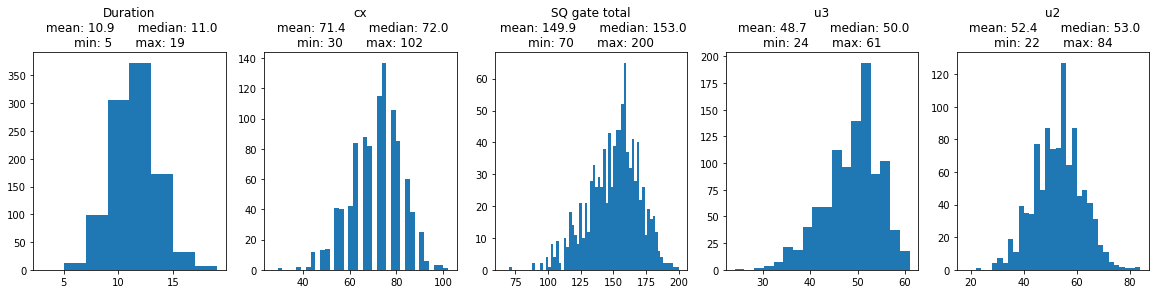}
    \caption{SABRE}
    \label{fig:6q_compiled_y_sabre}
    \end{subfigure}\\
    \begin{subfigure}{0.86\textwidth}
    \centering
    \includegraphics[width=\linewidth]{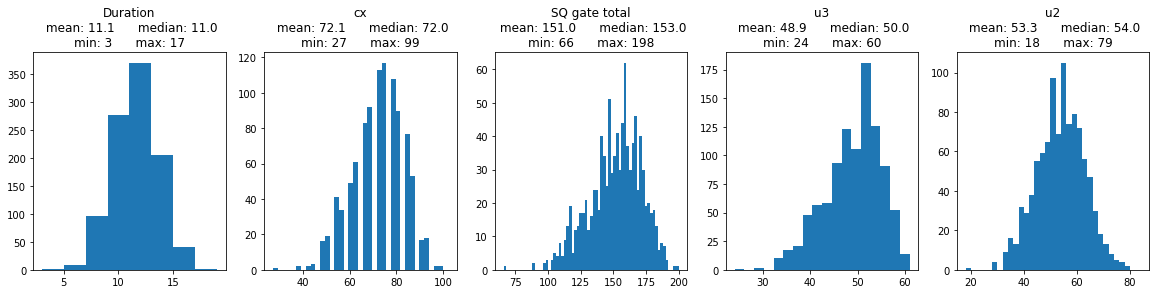}
    \caption{BIP-Layout}
    \label{fig:6q_compiled_y_bip_layout}
    \end{subfigure}\\
    \begin{subfigure}{0.86\textwidth}
    \centering
    \includegraphics[width=\linewidth]{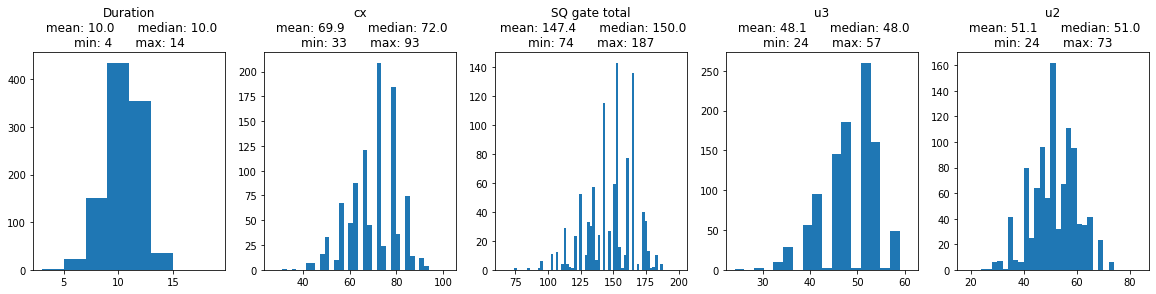}
    \caption{BIP-Routing}
    \label{fig:6q_compiled_y_bip_routing}
    \end{subfigure}\\
    \begin{subfigure}{0.86\textwidth}
    \centering
    \includegraphics[width=\linewidth]{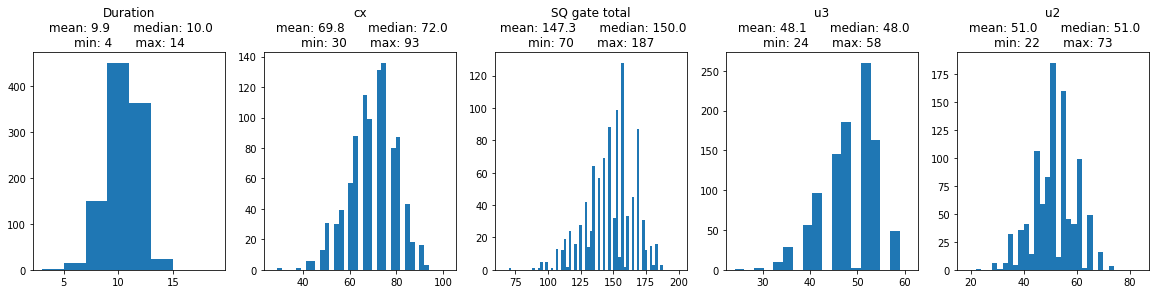}
    \caption{BIP-Constrained}
    \label{fig:6q_compiled_y_bip_constrained}
    \end{subfigure}
    \caption{Statistics for 6-qubit volume circuits compiled on the Y topology using different algorithms to solve the qubit allocation problem.}
    \label{fig:6q_compiled_y}
\end{figure}

\begin{figure}[tp]
    \centering
    \begin{subfigure}{0.86\textwidth}
    \centering
    \includegraphics[width=\linewidth]{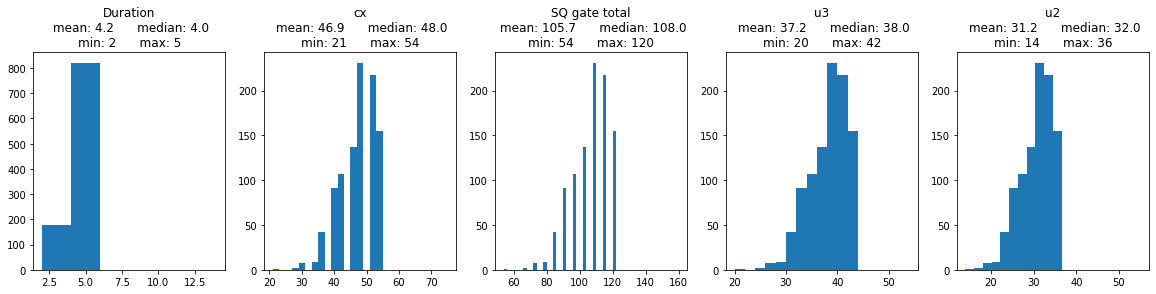}
    \caption{BIP}
    \label{fig:6q_compiled_grid_bip}
    \end{subfigure}\\
    \begin{subfigure}{0.86\textwidth}
    \centering
    \includegraphics[width=\linewidth]{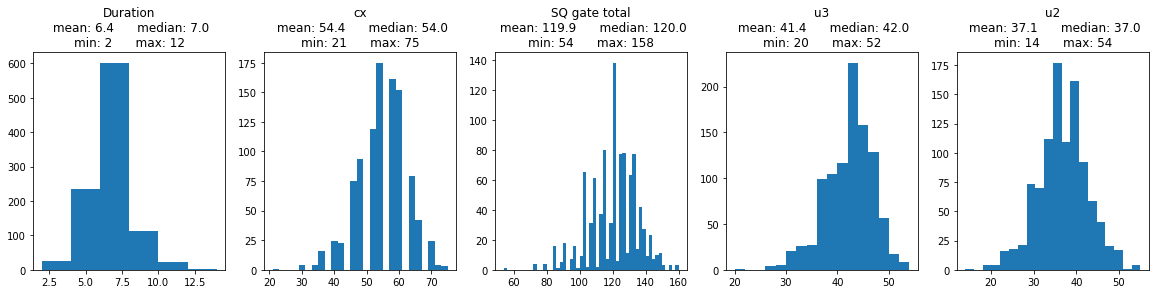}
    \caption{SABRE}
    \label{fig:6q_compiled_grid_sabre}
    \end{subfigure}\\
    \begin{subfigure}{0.86\textwidth}
    \centering
    \includegraphics[width=\linewidth]{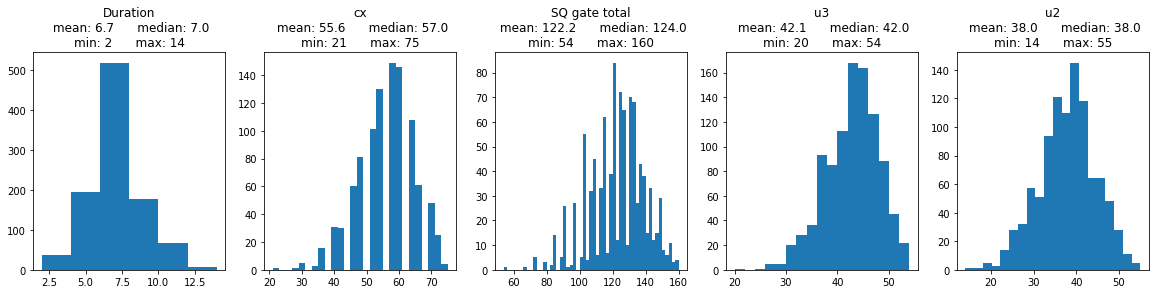}
    \caption{BIP-Layout}
    \label{fig:6q_compiled_grid_bip_layout}
    \end{subfigure}\\
    \begin{subfigure}{0.86\textwidth}
    \centering
    \includegraphics[width=\linewidth]{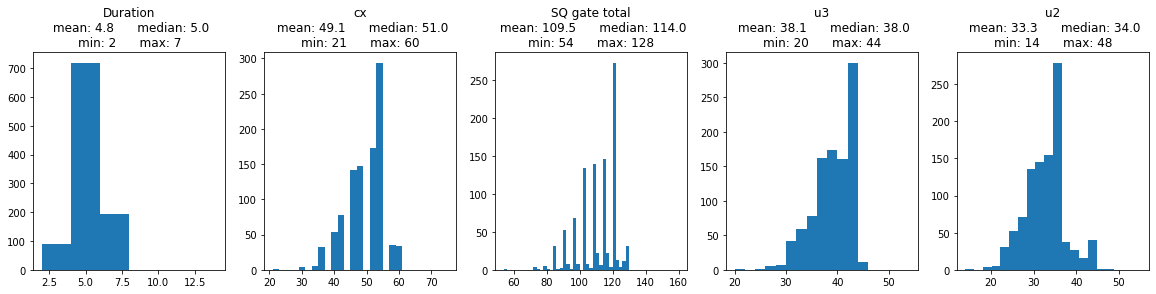}
    \caption{BIP-Routing}
    \label{fig:6q_compiled_grid_bip_routing}
    \end{subfigure}\\
    \begin{subfigure}{0.86\textwidth}
    \centering
    \includegraphics[width=\linewidth]{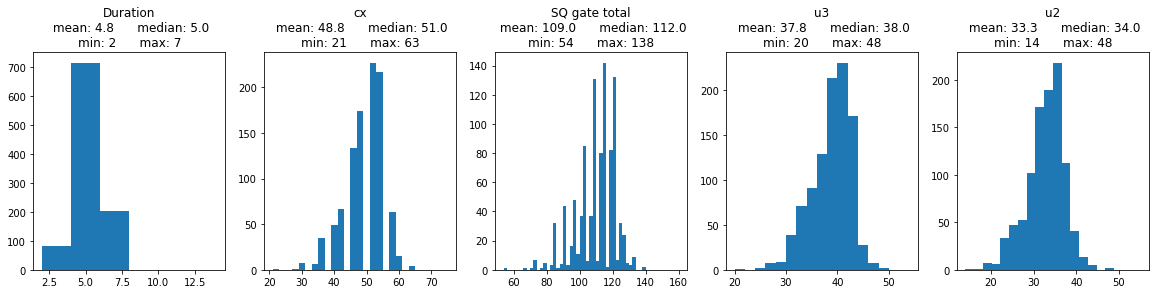}
    \caption{BIP-Constrained}
    \label{fig:6q_compiled_grid_bip_constrained}
    \end{subfigure}
    \caption{Statistics for 6-qubit volume circuits compiled on the grid topology using different algorithms to solve the qubit allocation problem. }
    \label{fig:6q_compiled_grid}
\end{figure}

Statistics on 1000 compiled circuits for all five algorithms on all three topologies are reported in Figures \ref{fig:6q_compiled_line}-\ref{fig:6q_compiled_grid}. The number of dummy time steps introduced between every two original circuit layers is set to 5; this ensures that every qubit permutation is reachable in between layers. For the grid topology, since we do not have access to hardware with the required connectivity, we report data as if the circuits were compiled for a stylized device that can execute each two-qubit gate in the same amount of time. Thus, results for the grid topology (Figure~\ref{fig:6q_compiled_grid}) are estimated, rather than measured.

In order to experimentally test the various transpilation algorithms we identify two subsets of six qubits on the 27 qubit processor $ibmq\_mumbai$ with a line (Q1-Q4-Q7-Q10-Q12-Q15) and a Y (Q1-Q4-Q7-Q10-Q12-Q6) topology. For these two topologies we randomly generate 300 (line) and 500 (Y) QV-circuits and pre-optimize them as explained above. Each circuit is executed 1000 times on the hardware and the heavy output probability of each individual circuit is estimated. The mean heavy output probabilities over all circuits for both topologies and the various methods are summarized in Table~\ref{tab:hop}. The reported error bars are given as the standard error of the sampling distribution $\sigma/\sqrt{n_c}$, with $\sigma$ as the standard deviation and $n_c$ as the number of executed circuits.


\begin{table}[htb]
    \centering
    \begin{tabular}{|l|c|c|}
    \hline
    Algorithm & HOP (line) & HOP (Y) \\
    \hline
    BIP & $0.6771  \pm 0.0022 $ & $0.6261 \pm 0.0019$  \\
    SABRE & $0.6506 \pm 0.0030$ & $0.6062 \pm 0.0020$ \\
    BIP-Layout & $0.6292 \pm 0.0031$ & $0.6000 \pm 0.0022$\\
    BIP-Routing & $0.6721 \pm 0.0023$ & $0.6221 \pm 0.0018$\\ 
    BIP-Constrained & $0.6710\pm0.0023$ & $0.6189\pm 0.0019$\\
    \hline
    \end{tabular}
    \caption{Heavy output probability measured on a line topology and a Y topology, using different algorithms to solve the qubit allocation problem.}
    \label{tab:hop}
\end{table}

The results in Table~\ref{tab:hop} indicate that BIP attains the largest HOP on the tested topologies. On the line topology, BIP-Routing and BIP-Constrained achieve similar values, lower than BIP by a small but significant margin, SABRE is in fourth place and BIP-Layout ranks last. These observations are consistent with Figure \ref{fig:6q_compiled_line}, where BIP attains shorter duration and smaller gate counts, suggesting that it produces circuits more resilient to noise. Another important observation is the fact that BIP-Constrained, while not as effective as BIP, is still more effective than SABRE: its median duration and median CNOT count are 10\% smaller, and it attain larger HOP. Thus, the restriction that the initial and final qubit permutation are the same is significant (compared to BIP, the median number of CNOTs with BIP-Constrained is 9\% larger), but the improvement of BIP over SABRE is not due only to the fact that BIP does not require the initial and final permutation to be the same. The situation is quite similar on the Y topology: the main difference is that BIP-Routing performs almost as well as BIP in this case, whereas the ranking of the remaining algorithms remains unchanged. The HOP values are consistent with the data in Figure \ref{fig:6q_compiled_y}. This indicates that on the Y topology, the routing component of the qubit assignment problem acquires even more importance, due to the increased potential for collisions at the central junction of the Y.  On the grid topology, even if we cannot test the circuits on physical hardware, the duration and CNOT count reported in Figure ~\ref{fig:6q_compiled_grid} indicate that the behavior of the five tested algorithms is consistent with the other topologies. On average, across the three topologies, BIP reduces the CNOT count by $\approx 10\%$ with respect to SABRE, and the duration of the circuit by $\approx 21\%$. These large reductions easily explain the improvement in the HOP recorded by BIP. The average CNOT count and duration are summarize in Table~\ref{tab:avg_6q}.

\begin{table}[tb]
    \centering
    \begin{tabular}{|l|c|c|c|c|c|c|}
    \hline
    & \multicolumn{2}{c|}{Line} & \multicolumn{2}{c|}{Y} & \multicolumn{2}{c|}{Grid} \\
    \cline{2-7}
    Algorithm & CNOT & Dur. & CNOT & Dur. & CNOT & Dur. \\
    \hline
    BIP     & 67.2 & 8.1 & 67.2 & 9.4 & 46.9 & 4.2 \\
    SABRE     &  73.8 & 9.7 & 71.4 & 10.9 & 54.0 & 6.4 \\
    BIP-Layout & 75.8 & 10.4 & 72.1 & 11.1 & 55.6 & 6.7 \\
    BIP-Routing & 70.5 & 8.8 & 69.9 & 10.0 & 49.1 & 4.8 \\
    BIP-Constrained & 70.5 & 8.7 & 69.8 & 9.9 & 48.8 & 4.8 \\
    \hline
    \end{tabular}
    \caption{Average CNOT count and duration for the 6-qubit quantum volume circuits.}
    \label{tab:avg_6q}
\end{table}

\begin{figure}[btp]
    \centering
    \begin{subfigure}{0.48\textwidth}
    \centering
    \includegraphics[width=\linewidth]{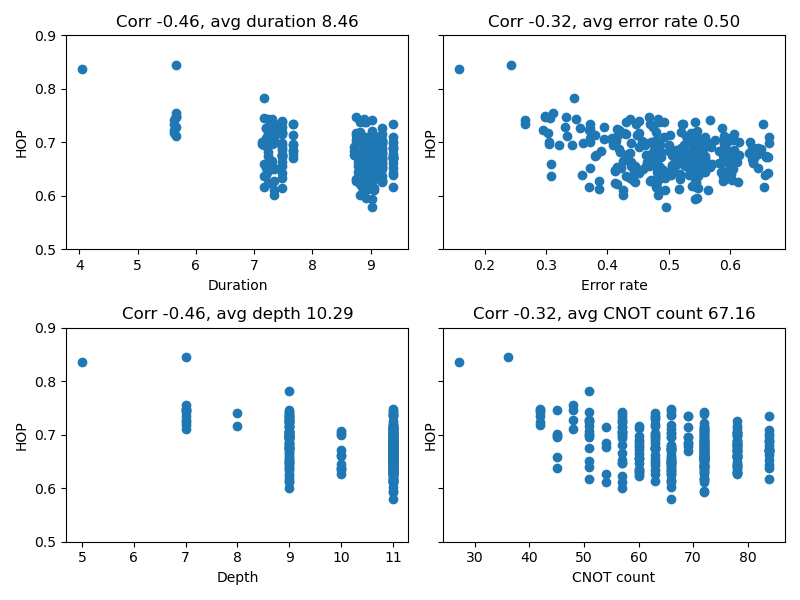}
    \caption{BIP}
    \label{fig:6q_exp_data_line_bip}
    \end{subfigure}%
    \begin{subfigure}{0.48\textwidth}
    \centering
    \includegraphics[width=\linewidth]{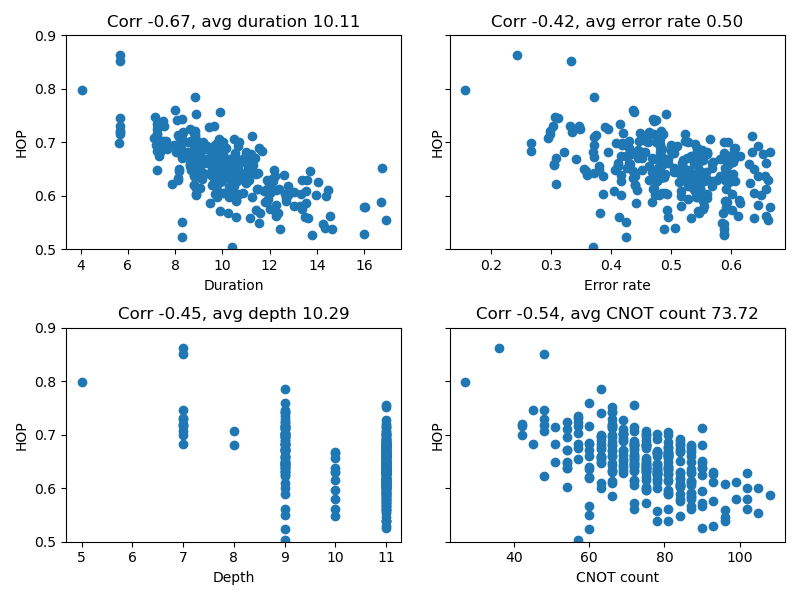}
    \caption{SABRE}
    \label{fig:6q_exp_data_line_sabre}
    \end{subfigure}\\
    \begin{subfigure}{0.48\textwidth}
    \centering
    \includegraphics[width=\linewidth]{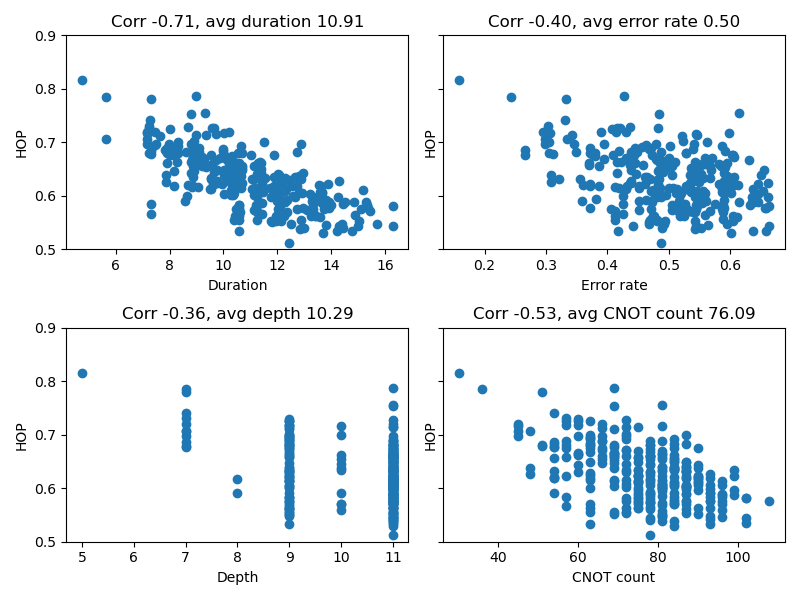}
    \caption{BIP-Layout}
    \label{fig:6q_exp_data_line_bip_layout}
    \end{subfigure}%
    \begin{subfigure}{0.48\textwidth}
    \centering
    \includegraphics[width=\linewidth]{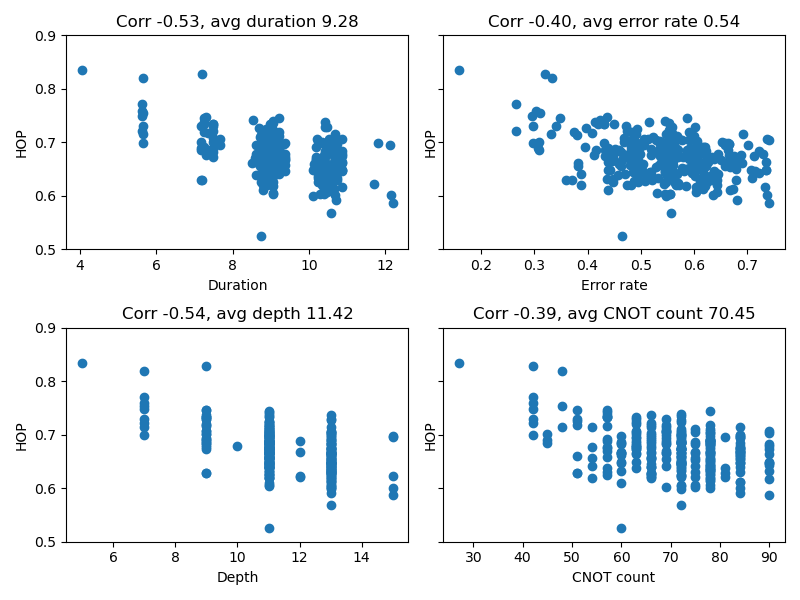}
    \caption{BIP-Routing}
    \label{fig:6q_exp_data_line_bip_routing}
    \end{subfigure}
    \caption{Scatter plot for 6-qubit volume circuits compiled on the line topology.}
    \label{fig:6q_exp_data_line}
\end{figure}

\begin{figure}[tp]
    \centering
    \begin{subfigure}{0.48\textwidth}
    \centering
    \includegraphics[width=\linewidth]{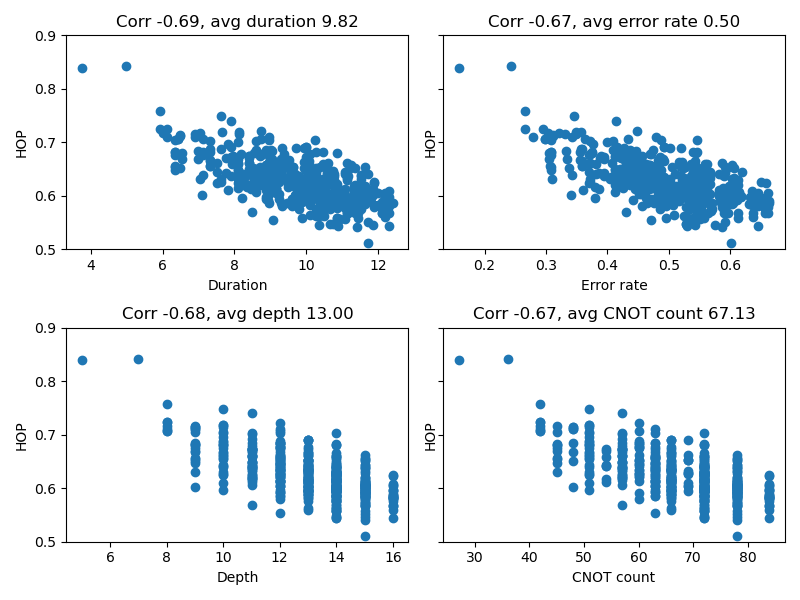}
    \caption{BIP}
    \label{fig:6q_exp_data_y_bip}
    \end{subfigure}%
    \begin{subfigure}{0.48\textwidth}
    \centering
    \includegraphics[width=\linewidth]{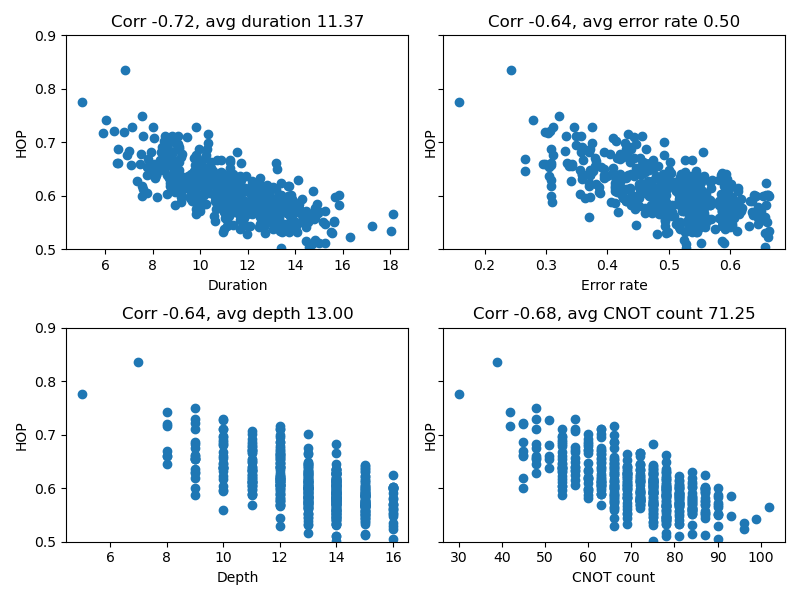}
    \caption{SABRE}
    \label{fig:6q_exp_data_y_sabre}
    \end{subfigure}\\
    \begin{subfigure}{0.48\textwidth}
    \centering
    \includegraphics[width=\linewidth]{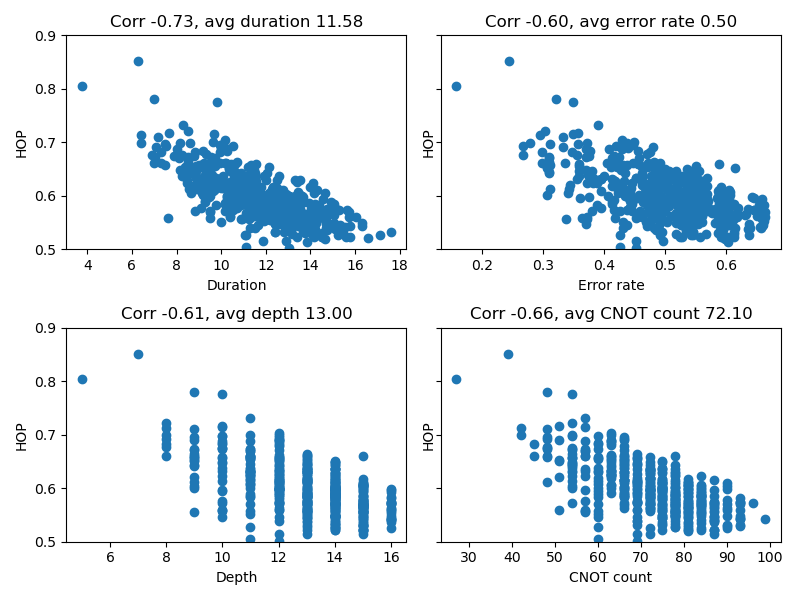}
    \caption{BIP-Layout}
    \label{fig:6q_exp_data_y_bip_layout}
    \end{subfigure}%
    \begin{subfigure}{0.48\textwidth}
    \centering
    \includegraphics[width=\linewidth]{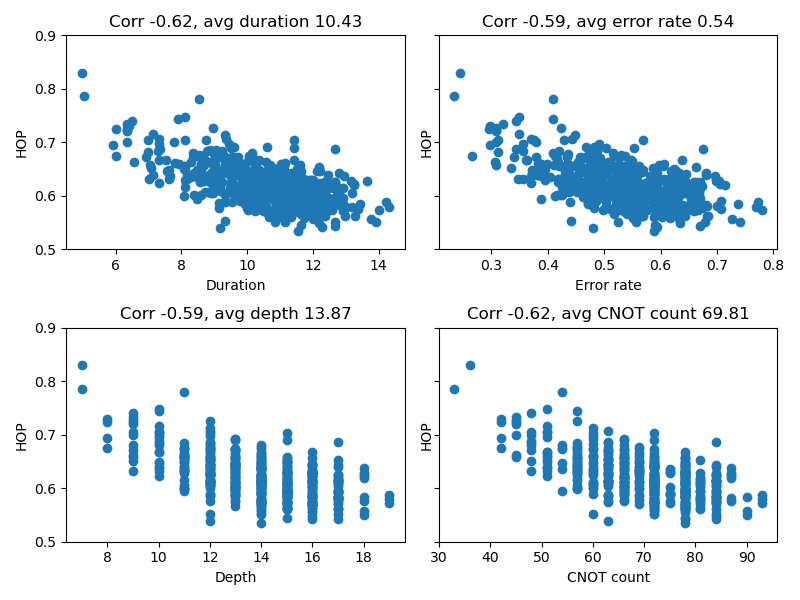}
    \caption{BIP-Routing}
    \label{fig:6q_exp_data_y_bip_routing}
    \end{subfigure}
    \caption{Scatter plots for 6-qubit volume circuits compiled on the Y topology.}
    \label{fig:6q_exp_data_y}
\end{figure}

We remark that in principle we would like to solve the qubit allocation problem in such a way as to maximize the resulting HOP, because this is the chosen measure of (experimental) quality of the circuit implementation. Since we do not know how to effectively model the HOP as an objective for the BIP, we instead optimize a proxy for the error rate and a measure of circuit depth, as explained in Section~\ref{s:objfun}. It is therefore important to test the correlation between the objective functions that we optimize, and the HOP; we additionally include the total circuit duration and CNOT count in the plots. In Figures~\ref{fig:6q_exp_data_line}-\ref{fig:6q_exp_data_y} we report scatter plots for duration, error rate, depth (these two are precisely the objective functions of Section~\ref{s:objfun}), and CNOT count, versus HOP. The plots also report the Pearson correlation coefficient. We only report four of the five algorithms because all plots are similar, and no further insight is gained from the missing plots. In these figures we observe that the duration of the circuit has the highest correlation coefficient with HOP; the objective functions used in the optimization, namely the error rate proxy and depth, exhibit mild to strong negative correlation with HOP. The CNOT count is similar to our proxy for the error rate: on this dataset, the Pearson correlation coefficient between error rate and CNOT count is between 0.84 and 0.99, depending on the algorithm used; this is expected, as the number of CNOT gates is the primary contributor to objective function \eqref{eq:objerrorrate}. Thus, minimizing the CNOT in addition to minimizing the error rate would be largely redundant.

We also experiment with a different characterization of the error rate, where the fidelity of each individual gate in the circuit is experimentally measured just before the execution of the circuit; this is contrast to our standard approach of using an average CNOT error rate, estimated from historical data, regardless of the specific gates to be executed on the device. We use the individual per-gate error rates to compute the total circuit error rate using the formula $1-\prod_{g \in G} (1- \epsilon_g)$, where $\epsilon_g$ is the error rate for gate $g \in G$; let us call this number ``measured error rate'', as opposed to the estimated error rate obtained from historical averages of the CNOT fidelity. On average, the Pearson correlation coefficient between the measured error rate and the HOP is 9\% higher than the correlation coefficient between the estimated error rate and the HOP; this indicates that it is likely to be a better predictor in practice. While the measured error rate can be used as the objective function, simply by appropriately choosing the data in the formula \eqref{eq:objerrorrate}, it requires running a procedure to estimate the error rate of individual gates just before the circuit compilation and execution, because this data changes over time. If such data is not available or is considered too expensive to compute, the approximation obtained by assigning the same basis fidelity to all edges (we used $\beta_{ij} = 0.9936$ in our experiments) yields a useful objective function already.

In summary, while not perfect predictors of the HOP, the objective functions used in our formulation are shown to drive the optimization in the desired direction, leading to larger HOP. To improve the correlation of the objective function with the HOP, it is possible to employ more accurate data obtained by measuring gate error rates on the device just before execution of the circuits; otherwise, a reasonable estimate of the basis fidelity is sufficient.

\subsubsection{8-qubit volume circuits}
We test the proposed formulation on 1000 8-qubit volume circuits. For these larger circuits we are unable to solve the BIP to proven optimality in a short time: this is because not only the logical circuit and the hardware graph are larger, but also we need to introduce more dummy time steps, in principle. To keep the size of the model under control, the number of dummy time steps introduced between every two original circuit layers is set to 4. This implies that not all qubit permutations are reachable in between layers, but in practice, it is unlikely that the optimal solution contains long sequences of SWAPs in between layers of the original circuit; thus, we do not expect this restriction to severely affect the quality of the returned solution. 

\begin{figure}[tp]
    \centering
    \begin{subfigure}{0.9\textwidth}
    \centering
    \includegraphics[width=\linewidth]{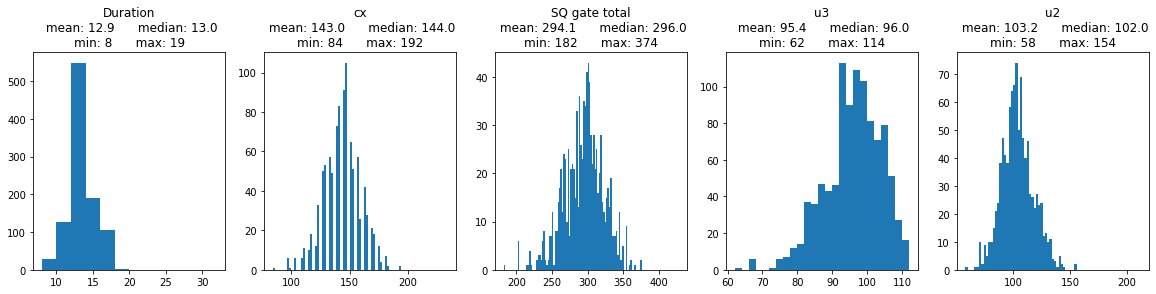}
    \caption{BIP}
    \label{fig:8q_compiled_line_bip}
    \end{subfigure}\\
    \begin{subfigure}{0.9\textwidth}
    \centering
    \includegraphics[width=\linewidth]{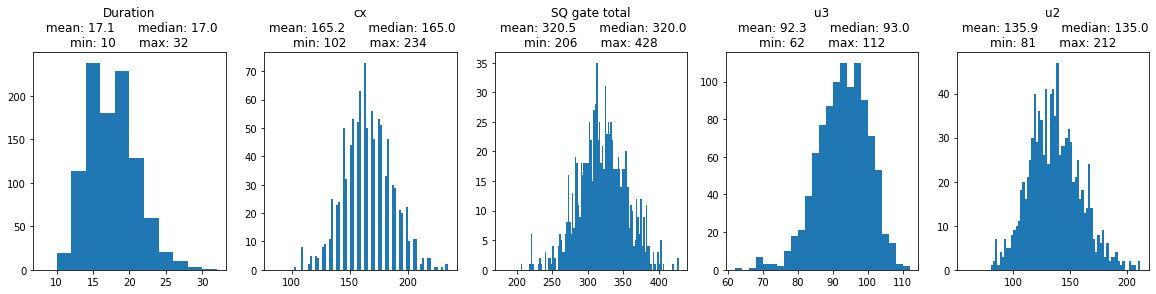}
    \caption{SABRE}
    \label{fig:8q_compiled_line_sabre}
    \end{subfigure}
    \caption{Statistics for 8-qubit volume circuits compiled on the line topology using BIP and SABRE to solve the qubit allocation problem. }
    \label{fig:8q_compiled_line}
\end{figure}

\begin{figure}[tp]
    \centering
    \begin{subfigure}{0.9\textwidth}
    \centering
    \includegraphics[width=\linewidth]{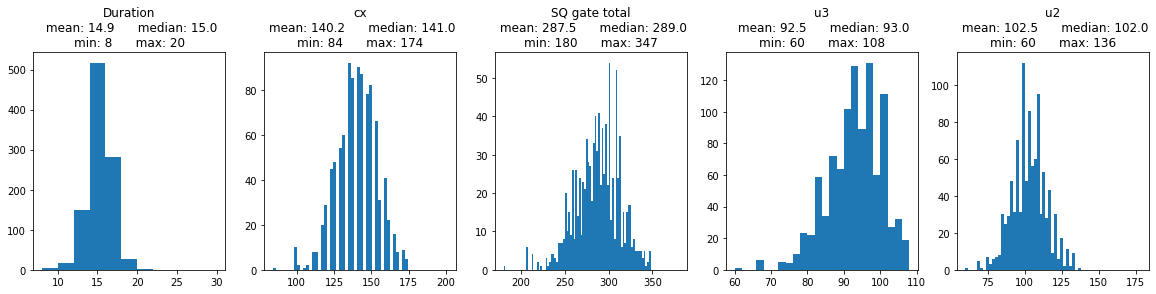}
    \caption{BIP}
    \label{fig:8q_compiled_y_bip}
    \end{subfigure}\\
    \begin{subfigure}{0.9\textwidth}
    \centering
    \includegraphics[width=\linewidth]{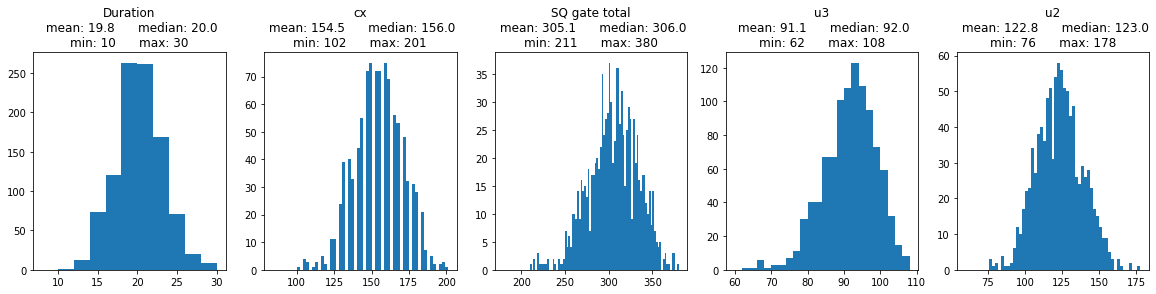}
    \caption{SABRE}
    \label{fig:8q_compiled_y_sabre}
    \end{subfigure}
    \caption{Statistics for 8-qubit volume circuits compiled on the Y topology using BIP and SABRE to solve the qubit allocation problem. }
    \label{fig:8q_compiled_y}
\end{figure}

\begin{figure}[tp]
    \centering
    \begin{subfigure}{0.9\textwidth}
    \centering
    \includegraphics[width=\linewidth]{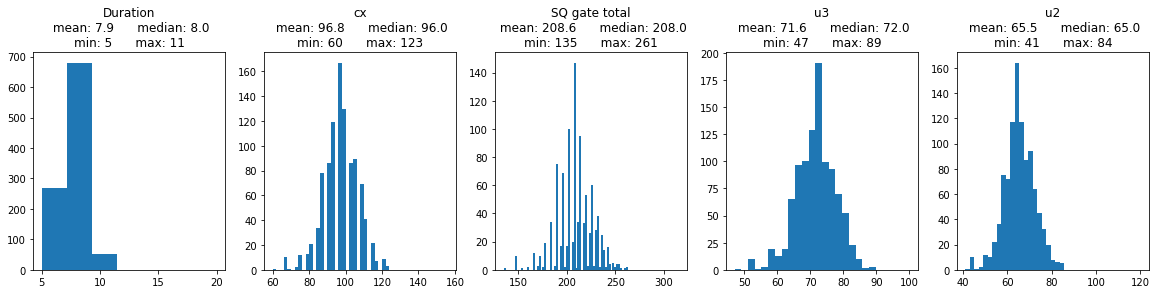}
    \caption{BIP}
    \label{fig:8q_compiled_grid_bip}
    \end{subfigure}\\
    \begin{subfigure}{0.9\textwidth}
    \centering
    \includegraphics[width=\linewidth]{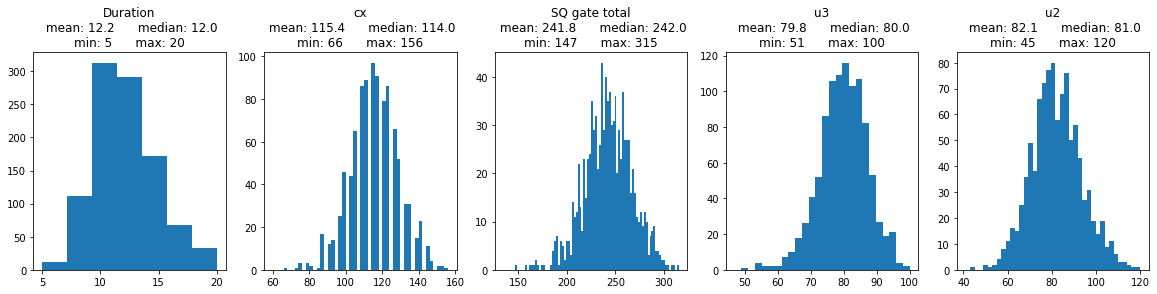}
    \caption{SABRE}
    \label{fig:8q_compiled_grid_sabre}
    \end{subfigure}
    \caption{Statistics for 8-qubit volume circuits compiled on the grid topology using BIP and SABRE to solve the qubit allocation problem. }
    \label{fig:8q_compiled_grid}
\end{figure}

Our computational experience indicates that solving these circuits to optimality typically requires 3-4 hours, using CPLEX with 10 threads. Since this is too long for practical purposes, we instead use CPLEX purely as a heuristic: we limit its CPU time to 9 minutes to minimize the error rate, and 1 additional minute to minimize depth after fixing the maximum error rate; the ``MIP emphasis'' parameter of CPLEX is set to 4, increasing the effort devoted to finding feasible solutions rather than proving optimality. With these settings, our goal is to show that our mathematical model for the qubit assignment problem can be useful in guiding the search toward good circuit implementations, even if we cannot determine a globally optimal one. In this section we only report results to compare BIP with SABRE; although we tested the other BIP variants, because the models are not solved to optimality we cannot draw any conclusions on the relative impact of the various components of BIP (i.e., the routing and layout). The data is summarized in Figures~\ref{fig:8q_compiled_line}-\ref{fig:8q_compiled_grid}. We remark that the data reported in this section is estimated with the simplifying assumption that all two-qubit gates have the same duration; this is due to the fact that we are not executing these circuits on real hardware, and is similar to what we did for the grid topology in the previous section.

Even when used within a heuristic search, the proposed BIP models finds circuits of much higher quality than SABRE in 10 minutes. The average CNOT count decreases by $\approx 14\%$ on the line topology, $\approx 9\%$ on the Y topology, and $\approx 16\%$ on the grid topology. The (estimated) duration is also significantly reduced, e.g., by $25\%$ on the line topology. The average CNOT count and duration are reported in Table~\ref{tab:avg_8q}.

\begin{table}[tb]
    \centering
    \begin{tabular}{|l|c|c|c|c|c|c|}
    \hline
    & \multicolumn{2}{c|}{Line} & \multicolumn{2}{c|}{Y} & \multicolumn{2}{c|}{Grid} \\
    \cline{2-7}
    Algorithm & CNOT & Dur. & CNOT & Dur. & CNOT & Dur. \\
    \hline
    BIP     & 143.0 & 12.9 & 140.2 & 14.9 & 96.8 & 7.9 \\
    SABRE     &  165.2 & 17.1 & 154.5 & 19.8 & 115.4 & 12.2 \\
    \hline
    \end{tabular}
    \caption{Average CNOT count and duration for the 8-qubit quantum volume circuits.}
    \label{tab:avg_8q}
\end{table}

\subsection{Beyond quantum volume circuits}
To prove the usefulness of our optimization model in a different context, we analyze its performance on a class of Clifford circuits, representing a toy model of Hamiltonian evolution, presented in \cite{bravyi2021clifford}. These circuits alternate between layers of CZ gates and layers of Hadamard gates, where the two-qubit connectivity lives on the graph $H = (V, E)$. Clifford circuits can be synthesized with many different techniques, and we use the optimized circuits produced with the algorithm of \cite{bravyi2021clifford}. Crucially, the optimized version of these Hamiltonian evolution circuits does not satisfy the connectivity constraints of the edge set $E$, in general, even though the optimization reduces the CNOT by a large amount. We use our BIP to solve the qubit assignment problem and map the optimized Clifford circuits back to the hardware topology $H$. This is a realistic usage scenario that goes beyond the QV circuits studied in the rest of this paper. Following \cite{bravyi2021clifford}, we only look at the CNOT count of the resulting circuits; we compare it with the CNOT count of the original Hamiltonian evolution circuits (i.e., before they are synthesized with the algorithm of \cite{bravyi2021clifford}), which already satisfy the hardware topology, and with the CNOT count obtained by using SABRE to solve the qubit assignment problem.

We compile all Hamiltonian evolution circuits of \cite{bravyi2021clifford} with at most $9$ qubits; this includes 49 circuits and 5 different topologies. We run BIP as a heuristic, with a limit of 10 minutes per circuit. The average CNOT counts are reported in Table~\ref{tab:clifford_cnot}. The CNOT count of the original circuits is much larger than for the optimized circuits: even after mapping the optimized circuits to the hardware topology (with either BIP or SABRE), we achieve a significant gain. On these circuits, BIP reduces the number of CNOTs by $\approx 11\%$ on average compared to SABRE, in line with the experiments on the QV circuits. Note that our approach to merge any single-qubit gate into neighboring two-qubit gates, and treat the entire circuit as a sequence of two-qubit gates, may affect some of the circuit optimizations applied by Qiskit. Nonetheless, applying BIP with the same settings as discussed in previous sections yields an advantage over SABRE. 

\begin{table}[htb]
    \centering
    \begin{tabular}{|l|c|c|c|}
    \hline
    &  Original & BIP & SABRE \\
    \hline
    CNOT count & 42.53 (4.83) & 15.96 (1.45) & 17.94 (1.85) \\
    \hline
    \end{tabular}
    \caption{Average CNOT counts; the standard error is given in brackets.}
    \label{tab:clifford_cnot}
\end{table}

\section{Conclusions}
\label{s:conclusions}
This paper introduces a mathematical optimization model for the qubit assignment problem,  and uses it to solve to global optimality instances of the problem derived from quantum volume circuits, that are designed to challenge existing hardware. In constrast to other papers in the open literature, our formulation is very general and can be applied to arbitrary hardware topologies. An important feature of the formulation is that it allows merging two-qubit gates with adjacent SWAPs at lower cost than implementing the gates and the SWAPs independently, reflecting the reality of the hardware. Our experimental evaluation, that employs a mix of simulated results on stylized devices and experiments on IBM's quantum hardware, leads to several important observations:
\begin{itemize}
    \item Two of the objective functions used in our model, namely, a proxy for the error rate and the depth of the circuit, exhibit mild to strong correlation with the heavy output probability of the circuits when executed on hardware. Optimizing these objective functions yields circuits with better fidelity.
    \item The depth and error rate can be optimized at the same time, with no trade-offs, for all circuits tested in this paper. While this trend is not expected to continue for all circuit types and sizes, it suggests that we can drive two important quality measures of a circuit in the desired direction at the same time, e.g., by optimizing a weighted combination of the two objectives.
    \item Reducing cross-talk, while an achievable goal in practice, conflicts with the optimization of error rate and depth, so that the trade-off between objectives should be carefully evaluated if we aim to reduce cross-talk.
    \item The default Qiskit transpiler, SABRE, produces reasonably good circuits, but is far from attaining the global optimum in several important metrics, e.g., total duration of the circuit and CNOT count: on average, the globally optimal solution produced by our mathematical optimization formulation have $\approx 20\%$ shorter duration and $\approx 10\%$ lower CNOT count. This reduction persists even when the mathematical optimization formulation is used to drive a heuristic search with an integer programming solver, i.e., it holds even if we do not solve the formulation to global optimality. On a set of toy Hamiltonian evolution circuits taken from \cite{bravyi2021clifford}, the mathematical optimization formulation reduces the CNOT count by $\approx 11\%$ on average.
    \item Among the components of the qubit assignment problem, i.e., the initial qubit layout and the qubit routing, determining an optimal routing is significantly more impactful on the final circuit fidelity.
    \item Restricting the initial and final qubit permutation of the circuit to be the same has a small but noticeble impact on the circuit fidelity, and can increase the circuit duration by $\approx 5\%$. The pros and cons of its adoption as a design choice for circuit implementation algorithms should be properly evaluated in light of this observation.
\end{itemize}
It is important to note that SABRE is significantly faster than our integer programming approach: already for 8-qubit volume circuits, our model requires a few minutes of CPU time to provide solutions that improve over SABRE. Thus, the proposed model is not suitable to solve large instances of the qubit assignment problem, and may be too slow for many practical applications. However, in cases where it is important to optimize the use of quantum resources as much as possible, this paper shows that mathematical optimization is a viable approach. Furthermore, our investigation provides insight on what are good metrics to optimize, and serves as a starting point for the development of mathematical optimization heuristics that can tackle larger circuits.

\bibliographystyle{plainurl}
\bibliography{compiler}

\appendix

\section{Additional figures}

\begin{figure}
    \centering
    \begin{subfigure}{0.5\textwidth}
    \centering
    \includegraphics[width=0.9\linewidth]{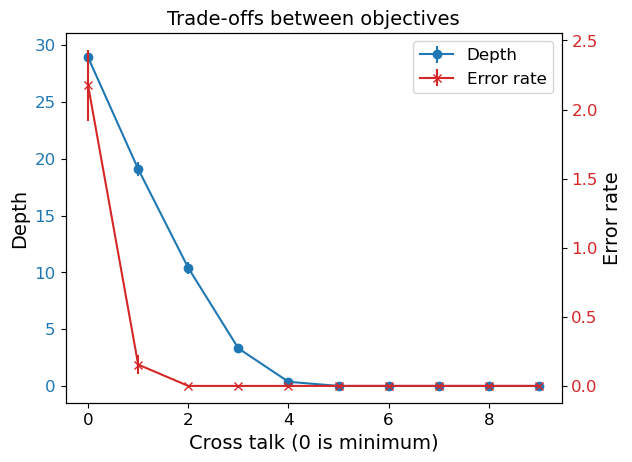}
    \caption{Cross talk vs.\ depth and error rate}
    \label{fig:pareto_cd_line}
    \end{subfigure}%
    \begin{subfigure}{0.5\textwidth}
    \centering
    \includegraphics[width=.9\linewidth]{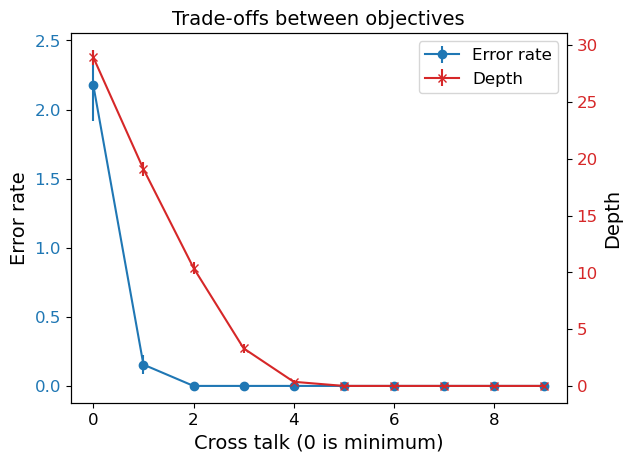}
    \caption{Cross talk vs.\ error rate and depth}
    \label{fig:pareto_ce_line}
    \end{subfigure}\\
    \begin{subfigure}{0.5\textwidth}
    \centering
    \includegraphics[width=.9\linewidth]{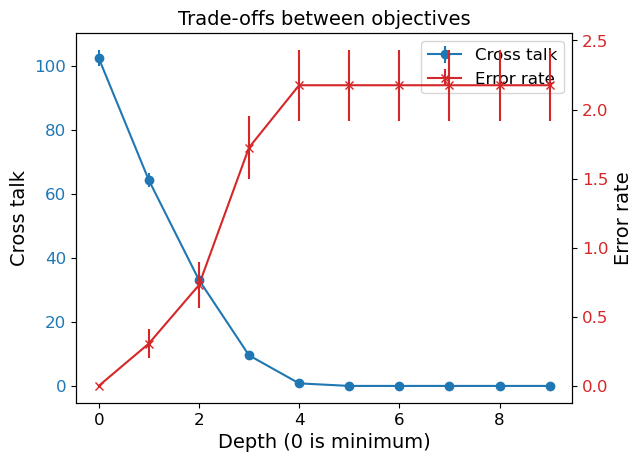}
    \caption{Depth vs.\ cross talk and error rate}
    \label{fig:pareto_dc_line}
    \end{subfigure}%
    \begin{subfigure}{0.5\textwidth}
    \centering
    \includegraphics[width=.9\linewidth]{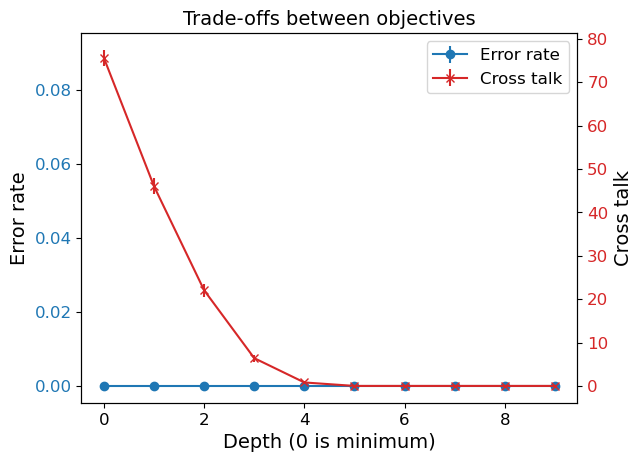}
    \caption{Depth vs.\ error rate and cross talk}
    \label{fig:pareto_de_line}
    \end{subfigure}\\
    \begin{subfigure}{0.5\textwidth}
    \centering
    \includegraphics[width=.9\linewidth]{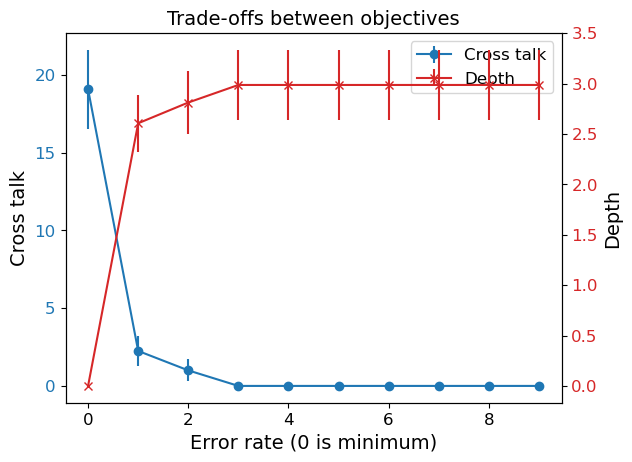}
    \caption{Error rate vs.\ cross talk and depth}
    \label{fig:pareto_ec_line}
    \end{subfigure}%
    \begin{subfigure}{0.5\textwidth}
    \centering
    \includegraphics[width=.9\linewidth]{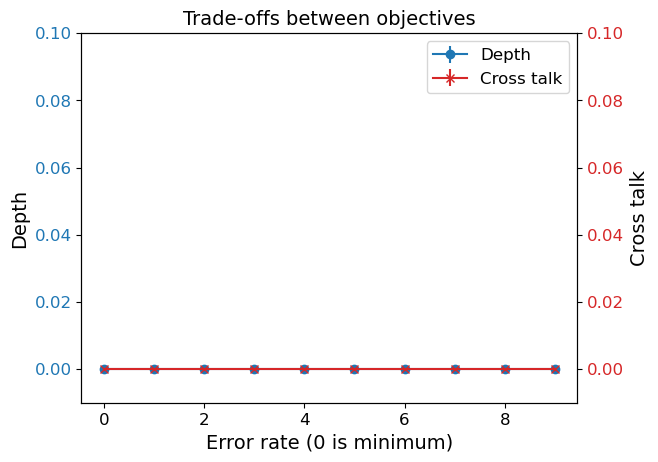}
    \caption{Error rate vs.\ depth and cross talk}
    \label{fig:pareto_ed_line}
    \end{subfigure}
    \caption{Pareto curves when optimizing three objectives in different orders, on a line topology. The $x$ axis contains the first objective to be optimized, with 0 indicating the global optimum. The left $y$ axis contains the second objective to be optimized. The right $y$ axis contains the third objective to be optimized. All $y$ values reported are relative increase with respect to the minimum. Vertical bars denote $\pm$ standard error.}
    \label{fig:pareto_line}
\end{figure}
\newpage
\begin{figure}
    \centering
    \begin{subfigure}{0.5\textwidth}
    \centering
    \includegraphics[width=0.9\linewidth]{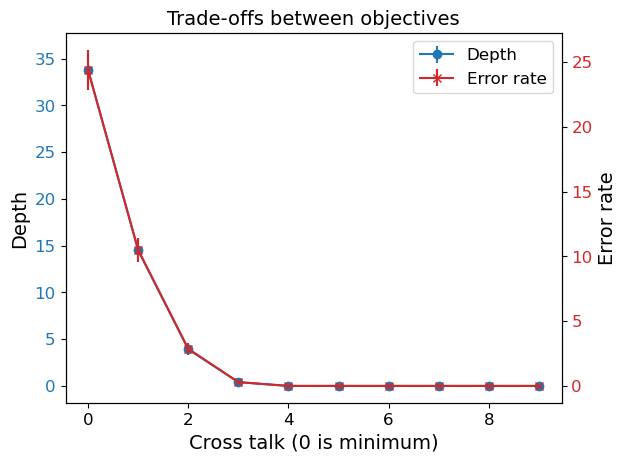}
    \caption{Cross talk vs.\ depth and error rate}
    \label{fig:pareto_cd_y}
    \end{subfigure}%
    \begin{subfigure}{0.5\textwidth}
    \centering
    \includegraphics[width=.9\linewidth]{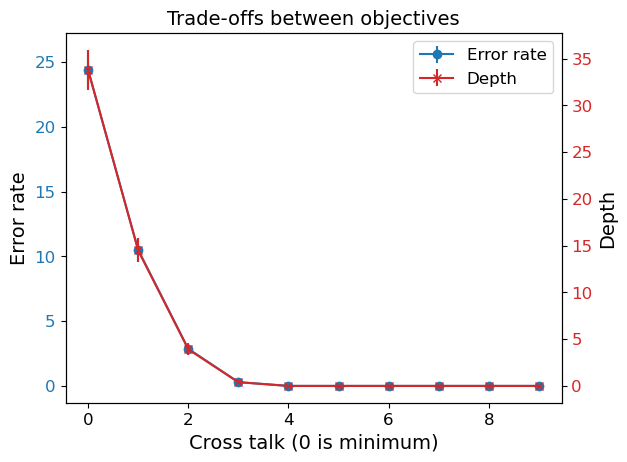}
    \caption{Cross talk vs.\ error rate and depth}
    \label{fig:pareto_ce_y}
    \end{subfigure}\\
    \begin{subfigure}{0.5\textwidth}
    \centering
    \includegraphics[width=.9\linewidth]{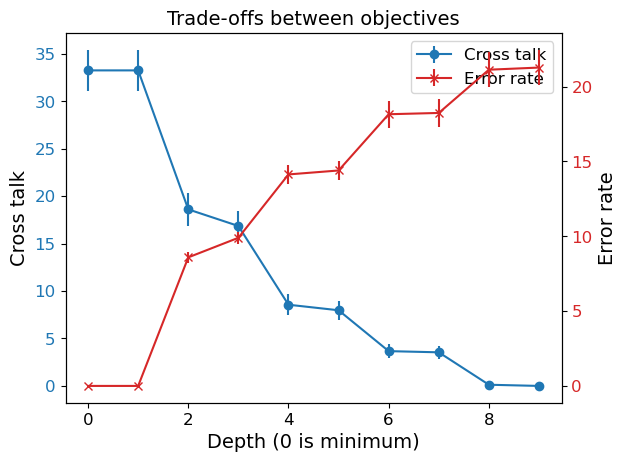}
    \caption{Depth vs.\ cross talk and error rate}
    \label{fig:pareto_dc_y}
    \end{subfigure}%
    \begin{subfigure}{0.5\textwidth}
    \centering
    \includegraphics[width=.9\linewidth]{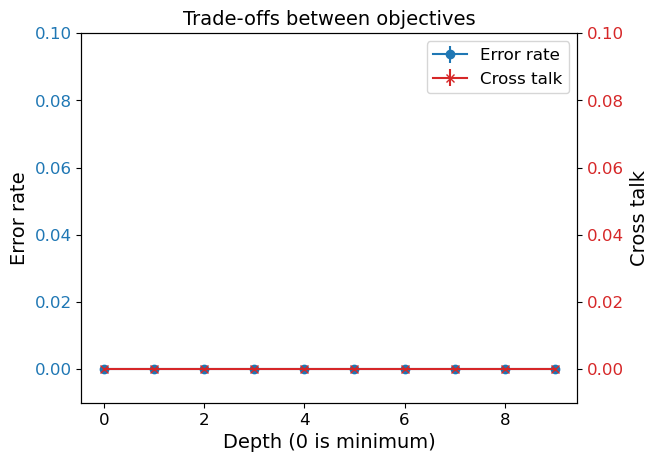}
    \caption{Depth vs.\ error rate and cross talk}
    \label{fig:pareto_de_y}
    \end{subfigure}\\
    \begin{subfigure}{0.5\textwidth}
    \centering
    \includegraphics[width=.9\linewidth]{figures/pareto_EC_y.png}
    \caption{Error rate vs.\ cross talk and depth}
    \label{fig:pareto_ec_y}
    \end{subfigure}%
    \begin{subfigure}{0.5\textwidth}
    \centering
    \includegraphics[width=.9\linewidth]{figures/pareto_ED_y.png}
    \caption{Error rate vs.\ depth and cross talk}
    \label{fig:pareto_ed_y}
    \end{subfigure}
    \caption{Pareto curves when optimizing three objectives in different orders, on a Y-shaped topology. The $x$ axis contains the first objective to be optimized, with 0 indicating the global optimum. The left $y$ axis contains the second objective to be optimized. The right $y$ axis contains the third objective to be optimized. All $y$ values reported are relative increase with respect to the minimum. Vertical bars denote $\pm$ standard error.}
    \label{fig:pareto_y}
\end{figure}

\begin{figure}
    \centering
    \begin{subfigure}{0.5\textwidth}
    \centering
    \includegraphics[width=0.9\linewidth]{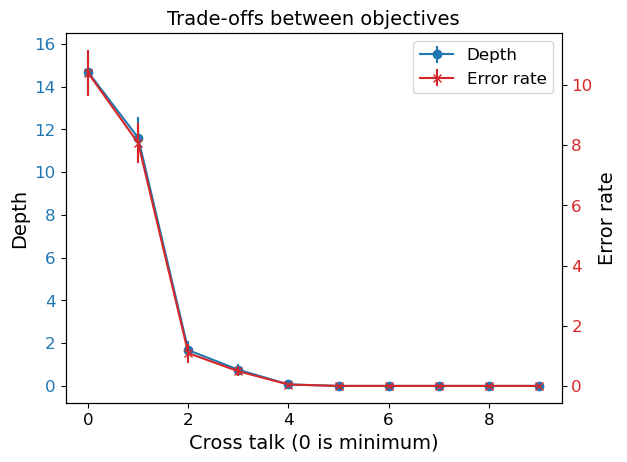}
    \caption{Cross talk vs.\ depth and error rate}
    \label{fig:pareto_cd_grid}
    \end{subfigure}%
    \begin{subfigure}{0.5\textwidth}
    \centering
    \includegraphics[width=.9\linewidth]{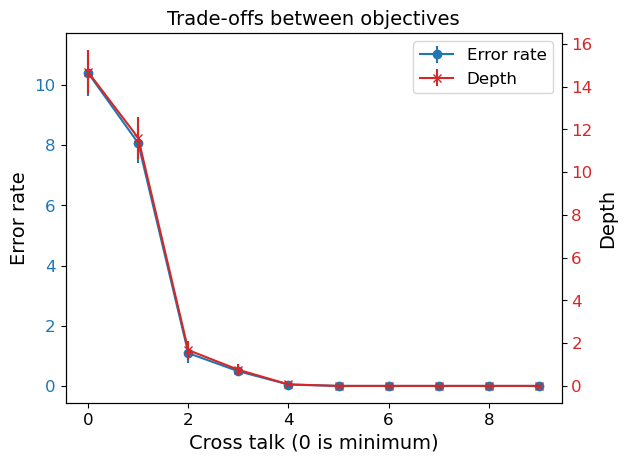}
    \caption{Cross talk vs.\ error rate and depth}
    \label{fig:pareto_ce_grid}
    \end{subfigure}\\
    \begin{subfigure}{0.5\textwidth}
    \centering
    \includegraphics[width=.9\linewidth]{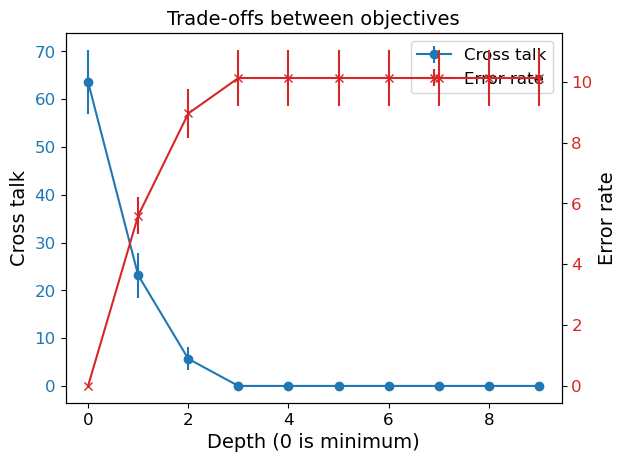}
    \caption{Depth vs.\ cross talk and error rate}
    \label{fig:pareto_dc_grid}
    \end{subfigure}%
    \begin{subfigure}{0.5\textwidth}
    \centering
    \includegraphics[width=.9\linewidth]{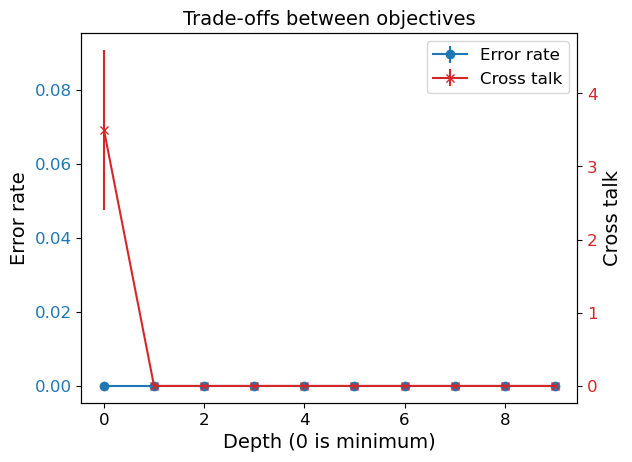}
    \caption{Depth vs.\ error rate and cross talk}
    \label{fig:pareto_de_grid}
    \end{subfigure}\\
    \begin{subfigure}{0.5\textwidth}
    \centering
    \includegraphics[width=.9\linewidth]{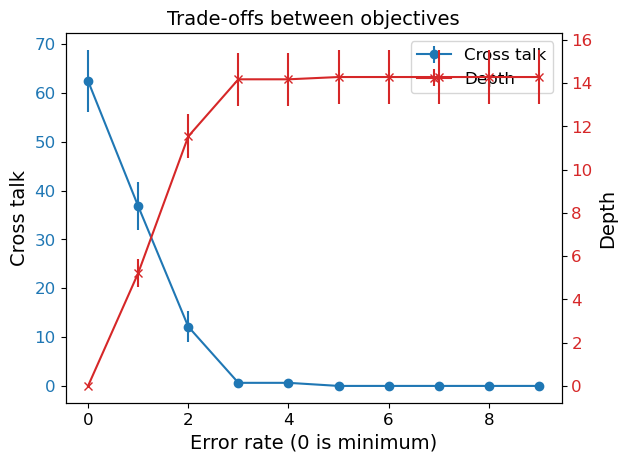}
    \caption{Error rate vs.\ cross talk and depth}
    \label{fig:pareto_ec_grid}
    \end{subfigure}%
    \begin{subfigure}{0.5\textwidth}
    \centering
    \includegraphics[width=.9\linewidth]{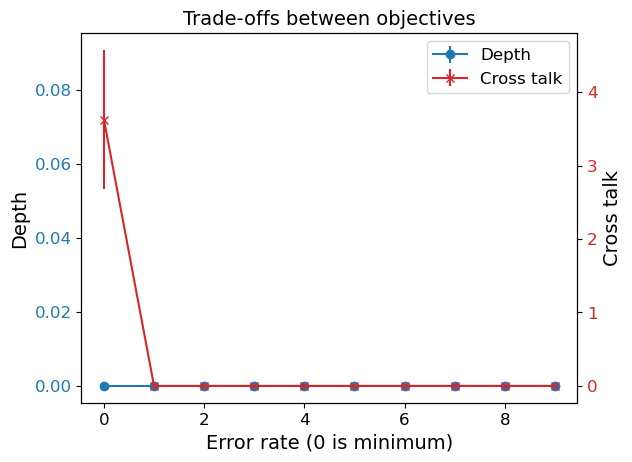}
    \caption{Error rate vs.\ depth and cross talk}
    \label{fig:pareto_ed_grid}
    \end{subfigure}
    \caption{Pareto curves when optimizing three objectives in different orders, on a grid topology. The $x$ axis contains the first objective to be optimized, with 0 indicating the global optimum. The left $y$ axis contains the second objective to be optimized. The right $y$ axis contains the third objective to be optimized. All $y$ values reported are relative increase with respect to the minimum. Vertical bars denote $\pm$ standard error.}
    \label{fig:pareto_grid}
\end{figure}

\end{document}